\documentclass[12pt,sort&compress]{article}

\usepackage[dvips]{color}
\usepackage{float}
\usepackage{amsmath}
\usepackage{graphicx}
\usepackage{tabularx}
\usepackage{subcaption}
\usepackage{cite}
\usepackage{amssymb}
\usepackage{mathtools}
\usepackage{lmodern}
\usepackage[sort&compress,numbers]{natbib}
\usepackage[toc,page]{appendix}
\usepackage{multicol}
\usepackage{multirow}

\usepackage{amsfonts}
\usepackage{tabularx,ragged2e}
\usepackage{xcolor}
\newcolumntype{C}{>{\Centering\arraybackslash}X} 
\usepackage[outercaption]{sidecap}    
\usepackage{hyperref}

\newcommand{\intB}{\int_{-\frac{\pi}{a}}^{\frac{\pi}{a}}}
\newcommand{\tauInv}{\tau^{-1}}
\newcommand{\taudInv}{(\tau d)^{-1}}
\newcommand{\sumd}{\sum_{i=1}^d}
\newcommand{\intP}{\int_{-\pi}^\pi}

\begin{document} 
\title{Voter model under stochastic resetting}
\date{}
\author{Pascal Grange\\
{\emph{Division of Natural and Applied Sciences}}\\
{\emph{and Zu Chongzhi Center for Mathematics and Computational Science}}\\
 Duke Kunshan University\\
8 Duke Avenue, Kunshan, 215316 Jiangsu, China\\
\normalsize{{\ttfamily{pascal.grange@dukekunshan.edu.cn}}}}
\maketitle

\begin{abstract}
The voter model is a toy model of consensus formation based on nearest-neighbor interactions. A voter sits at each vertex in a hypercubic lattice (of dimension $d$) and is in one of two possible opinion states.  The opinion state of each voter flips randomly, at a rate proportional to the fraction of the nearest neighbors that disagree with the voter. If the voters are initially independent and undecided, the model is known to lead to a consensus if and only if $d\leq 2$.  In this paper the model is subjected to stochastic resetting: the voters revert independently to their initial opinion according to a Poisson process of fixed intensity (the resetting rate). This resetting prescription induces kinetic equations for the average opinion state and for the two-point function of the model. For initial conditions consisting of undecided voters except for one decided voter at the origin, the one-point function evolves as the probability of presence of a diffusive random walker on the lattice, whose position is stochastically reset to the origin. The resetting prescription leads to a non-equilibrium steady state. For an initial state consisting of independent undecided voters, the density of domain walls in the steady state is expressed in closed form as a function of the resetting rate. This function is differentiable at zero  if and only if  $d\geq 5$.
\end{abstract}


\tableofcontents

\section{Introduction}

The voter model is a toy model of consensus formation in a population. Each vertex of a 
  hypercubic lattice of dimension $d$ is occupied by a voter, who is in one of two possible opinion states, say $-1$ and $+1$. Each voter flips its opinion state at a rate proportional to the fraction of its neighbors that disagree with it. In particular, if the system reached a configuration in which  all voters are in the same opinion state, the evolution stops (consensus is reached).  The following two  problems are particularly natural to consider:\\
 (I) For an initial condition with  undecided voters, except  one with a positive opinion, how does the opinion state of the initially decided voter evolve in time? Does it relax to the undecided state? To solve this problem, one needs to calculate the one-point function of the model. \\
 (II) For an initial condition consisting of independent undecided voters, is  consensus achieved in the long-time limit? To characterize how far from consensus the system is, one needs to estimate the density of domain walls.\\
 
 In a given configuration, domain walls are quasi-particles sitting at the bond between adjacent vertices occupied by voters in opposite opinion states. For example, in a configuration with independent undecided voters, the opinion states of the voters are independent centered Bernoulli random variables, and the density of domain walls is $1/2$ because two voters at adjacent vertices are in opposite opinion states with probability $1/2$.
 On the other hand, the density of domain walls is zero if there is consensus. It is enough to calculate the two-point function of the model to obtain the density of domain walls.\\

Quite remarkably, Problems (I) and (II) can be solved exactly 
 \cite{krapivsky1992kinetics,frachebourg1996exact}  in any dimension $d$ (for reviews, see \cite{liggett1985interacting,redner2019reality} and Chapter 8 of \cite{kineticView}).  The asymptotic behavior of the  one-point function implies that the opinion state of a single decided voter at the origin relaxes to the 
 undecided opinion state: the average opinion state at the origin goes to zero (as $d/(2\pi t)^{d/2}$) when time $t$ goes to infinity. For initial conditions corresponding to undecided voters, the system is translationally invariant, so the density of domain walls  depends only on time. 
 In \cite{frachebourg1996exact}, the large-time limit of the density  was calculated  and shown to be zero if and only if $d\geq 2$ (it decays to zero like the inverse square-root of time in dimension one, and like an inverse logarithm in dimension $2$). In higher dimension the limit is strictly positive and expressed in closed form in terms of integrals involving Bessel functions.\\

 In the voter model, individuals   
  are collaborative: they flip their opinion states based solely on the opinion states  of their neighbors. In this paper we modify the model to take into account a certain degree of stubbornness in the individuals. We do so by subjecting the voter model to stochastic resetting: we assume voters keep  memory of their initial opinion states and revert to it (independently) at random times. For each voter, these times are generated by a Poisson process of intensity $r$ (the quantity $r$ is called the resetting rate). The Poisson processes attached to different voters are independent. This kind of resetting prescription, according to which the degrees of freedom in the model are reset independently to their initial value, has been termed {\emph{local resetting}}
 in interacting particle systems  \cite{miron2021diffusion}.  More generally,
  stochastic resetting has received a considerable amount of attention 
  in out-of-equilibrium physics.  Indeed, stochastic resetting holds a system  away
from its long-time stationary state, leading to a non-equilibrium stationary state.
  The corresponding stationary distribution was first calculated
 exactly for a diffusive random walker \cite{evans2011diffusion,evans2011optimal}. Subsequent 
 developments yielded exact results on the non-equilibrium stationary state of a variety  
 of single-particle dynamics under resetting, including diffusion in potentials \cite{pal2015diffusion}, L\'evy flights \cite{kusmierz2014first} and active particles \cite{evans2018run,refractory,kumar2020active} (for a review, see \cite{topical,gupta2022stochastic}).
  Developments on systems with interacting degrees of freedom under local resetting include 
  models of binary aggregation \cite{grange2021aggregation}  and exclusion processes  \cite{miron2021diffusion,pelizzola2021simple}.
  In the other broad class of resetting dynamics ({\emph{global resetting}}), a system of interacting particles is reset to its initial configuration at Poisson-distributed time. In \cite{magoni2020ising}, the Ising model was subjected to stochastic resetting to a ferromagnetic state, which allowed to study the phases of the model (for a field-theoretic treatment of the Ising magnet reset to a paramagnetic state, see \cite{aron2020nonanalytic}). Global resetting has also been studied for exclusion processes \cite{basu2019symmetric,sadekar2020zero,karthika2020totally}, as well as predator-prey models \cite{quetzalcoatl2019predator,evans2022exactly,mercado2018lotka}, fluctuating interfaces \cite{gupta2014fluctuating,gupta2016resetting}, synchronization \cite{sarkar2022synchronization},  reaction-diffusion processes \cite{durang2014statistical}, glassy systems \cite{grange2020entropy} and zero-range processes \cite{ZRPSS,ZRPResetting}. 
  For a review of developments on interacting particle systems under stochastic resetting, see \cite{nagar2023stochastic} and references therein.\\

 Given the wide spectrum of systems for which non-equilibrium stationary states  under stochastic resetting can be worked out exactly, and given the simplicity of the dynamics of the voter model, we expect 
 to be able to work out one- and two-point functions in the voter model under stochastic resetting. 
 Intuitively, stochastic resetting should prevent the opinion state of a single initially decided voter to relax to the general undecided state. Moreover, if there is no consensus in the initial configuration, the dynamics of the voter model under resetting does not stop when consensus is reached. Indeed some of the voters will revert to their initial opinion state and break the consensus.  
    We are therefore led to the following two problems, the 
    analogs of Problems (I) and (II) under stochastic resetting:\\
(I$_{(r)}$)  For an initial condition with undecided voters, except one with a positive opinion at the site labeled $\mathbf{0}$ in the lattice, calculate 
 the average opinion state $S(\mathbf{0},\infty)$ of the initially decided voter in the steady state, as a function of the resetting rate $r$.\\
(II$_{(r)}$) For an initial condition consisting of independent undecided voters,  calculate the average density of domain walls $\varrho_\infty$ in the steady state.\\




The paper is organized as follows. In Section \ref{QOI} we define and study the model in dimension  one. We define the flipping rate of the voter model under stochastic resetting and   work out the evolution equation 
   of the average opinion-state profile  $S(x,t) =\langle s(x,t)\rangle$. We solve it in Fourier space.
   The evolution equation of the one-point function with a single initially decided is found to be closely related to the diffusion equation for a random  walker on the lattice, with stochastic resetting to the origin. 
 This yields the solution of Problem (I$_{(r)}$) in dimension one. We  work out the evolution equation  of the  two-point function, and notice that 
 the resetting prescription induces terms proportional to the {\emph{two-time}} two-point function $\langle s(x,t) s(y,0) \rangle$. However, at fixed $y$, this  function satisfies the same evolution equation as the average opinion state {\emph{with a single decided voter}}.
  The steady-state density of domain walls (the solution of Problem (II$_{(r)}$) in dimension one) is  obtained in closed form  by adapting the techniques
 of \cite{frachebourg1996exact}. In Section \ref{higherDim} we  generalize the results to higher dimension and work out the behavior at low resetting rate of the  density of domain walls in the steady state. Some of the mathematical details of the derivations are presented in appendices.

\section{The voter model under stochastic resetting in  dimension one}\label{QOI}

\subsection{Definitions and notations}
 The one-dimensional voter is defined as follows. At each vertex of a one-dimensional lattice sits a voter who can be in one of two opinion states, say $+1$ and $-1$.  
At time $t$, the opinion state of the voter at position  $x$ is denoted by $s(x,t)$, so that the configuration of the entire system  is given  by the collection
\begin{equation}\label{config}
\left\{s(x,t), x\in a\mathbb{Z} \right\} \in \left\{ -1, +1\right\}^{\mathbb{Z}},
\end{equation}
  where  $a$ denotes the lattice spacing. The opinion state of each voter flips at a rate proportional to the fraction of nearest neighbors that disagree with it. 
 Let us denote the  proportionality constant by 
$\tau^{-1}$ (the time $\tau$ sets the time scale of the system). If the voter at site $na$ is in a given opinion state, its opinion flips:\\
 $(i)$ at a rate of $\tau^{-1}$ if both of its neighbors are in the opposite opinion state,\\
 $(ii)$ at a rate of $\tau^{-1}/2$ if exactly only one of its neighbors is in the opposite 
opinion state,\\
 $(iii)$ at zero rate if none of its neighbors are in the opposite opinion state.\\
 All the possible cases are listed in Table \ref{tableSchematic}. The values of the flipping rate $W(na,t)$ displayed in the third column of the table are reproduced by the formula
\begin{equation}\label{WDef}
 W( na, t):=\frac{\tau^{-1}}{2}\left[ 1- \frac{1}{2}s(na,t)\left( s( (n+1)a,t) +  s((n-1)a,t) \right)  \right],\;\;\;\;\;\;\;\;\;\;(n\in \mathbb{Z}). 
\end{equation}
 Indeed the expression between square brackets takes the values $2,1,0$ in the cases $(i),(ii),(iii)$ respectively, because of the constraint $s(na,t)^2=1$.
The voters are cooperative in the sense that they change their opinion state based on the opinion states of their neighbors, without taking into account their  own initial opinion state. We will refer to the model with flipping rates $W$ as the ordinary voter model (solved in  \cite{krapivsky1992kinetics,frachebourg1996exact}).\\

\begin{table}
\begin{center}
\begin{tabular}{ | p{8cm}|p{2cm}|c|c|c|}
\hline
 $[s((n-1)a,t),s( na,t),s( (n+1)a,t)  ]$ & $s(na,0)$ & $W(na,t)$ & $R(na,t)$ & $w(na,t)$ \\
\hline
$[+1, +1, +1 ]$ & $+1$ & $0$ & $0$ & $0$ \\
$[+1, +1, -1 ]$ & $+1$ & $\frac{\tau^{-1}}{2}$ & $0$ & $\frac{\tau^{-1}}{2}$ \\
$[-1, +1, +1 ]$ & $+1$ & $\frac{\tau^{-1}}{2}$ & $0$ & $\frac{\tau^{-1}}{2}$ \\
$[-1, +1, -1 ]$ & $+1$ & $\tau^{-1}$ & $0$ & $\tau^{-1}$ \\
\hline
$[+1, -1, +1 ]$ & $-1$ & $\tau^{-1}$ & $0$ & $\tau^{-1}$ \\
$[+1, -1, -1 ]$ & $-1$ & $\frac{\tau^{-1}}{2}$ & $0$ & $\frac{\tau^{-1}}{2}$ \\
$[-1, -1, +1 ]$ & $-1$ & $\frac{\tau^{-1}}{2}$ & $0$ & $\frac{\tau^{-1}}{2}$ \\
$[-1, -1, -1 ]$ & $-1$ & $0$ & $0$ & $0$ \\
\hline
$[+1, +1, +1 ]$ & $-1$ & $0$ & $r$ & $r$ \\
$[+1, +1, -1 ]$ & $-1$ & $\frac{\tau^{-1}}{2}$ & $r$ & $\frac{\tau^{-1}}{2} +r$ \\
$[-1, +1, +1 ]$ & $-1$ & $\frac{\tau^{-1}}{2}$ & $r$ & $\frac{\tau^{-1}}{2}+r$ \\
$[-1, +1, -1 ]$ & $-1$ & $\tau^{-1}$ & $r$ & $\tau^{-1} + r$ \\
\hline
$[+1, -1, +1 ]$ & $+1$ & $\tau^{-1}$ & $r$ & $\tau^{-1}+r$ \\
$[+1, -1, -1 ]$ & $+1$ & $\frac{\tau^{-1}}{2}$ & $r$ & $\frac{\tau^{-1}}{2}+r$ \\
$[-1, -1, +1 ]$ & $+1$ & $\frac{\tau^{-1}}{2}$ & $r$ & $\frac{\tau^{-1}}{2}+r$ \\
$[-1, -1, -1 ]$ & $+1$ & $0$ & $r$ & $r$ \\
\hline
\end{tabular}
\end{center}
\caption{{\bf{The values of the flipping rate  at site  $na$ and time $t$ in each of the possible 
 configurations of the four opinion states $s((n-1)a,t),s( na,t),s( (n+1)a,t) $ and $s(na,0)$.}} The third column contains the flipping rate of the ordinary voter model, which depends only on the opinion states at time $t$ listed in the first column. The fourth column contains the additional flipping rate induced by resetting, which depends only on $s( na,t)$ and $s(na,0)$. The flipping rate in the voter model under stochastic resetting is $w(na,t)$, defined in Eq. (\ref{wDef}).}
\label{tableSchematic}
\end{table}

\begin{figure}
\begin{center}
 \includegraphics[width=17cm]{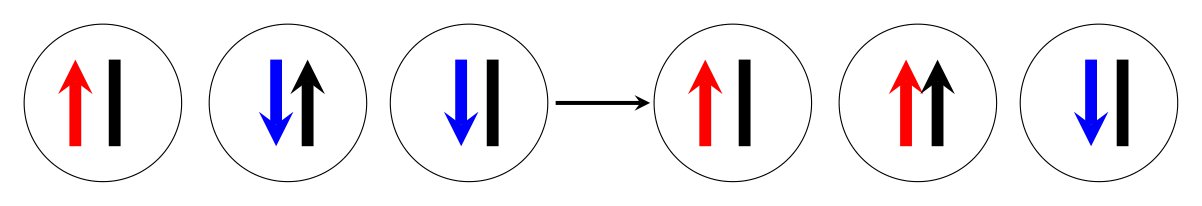}\\
\caption{{\bf{A flipping processe at site $na$ and time $t$.}} The coloured  arrows symbolize the current opinion states at sites $(n-1)a, na, (n+1)a$. The vertical black arrows symbolize the opinion state at site $na$ at time $0$. The  tips or the black arrows are not shown at sites  $(n-1)a, (n+1)a$, as the values of $s((n-1)a,0)$ and $s((n+1)a,0)$ are irrelevant to the flipping rate of $s(na,t)$. The opinion state at site $na$ is different from its initial value, and from the opinion state of one of the neighbors. The rate of the process is therefore $r+\tau^{-1}/2$.}
 \label{figSchematic}
\end{center}
\end{figure}

 Let us subject the voter model to stochastic resetting as follows. 
 We assume each  voter reverts to its initial opinion state at Poisson-distributed times.  A  Poisson process of intensity denoted by $r$ (the resetting rate) is attached to the site labeled $na$. It generates resetting times. At each resetting time, the variable $s(na,t)$ is reset to its initial value $s(na,0)$, if it differs from this initial value (otherwise it does not change).  The Poisson processes attached to different  sites are  independent. Hence, between times $t$ and $t+dt$, 
  the opinion state $s(na,t)$ reverts to $s(na,0)$ with probability $rdt$ if $s(na,t)\neq s(na,0)$.
This resetting prescription induces an additional flipping rate at site $na$ and time $t$, denoted by $R(na,t)$. The possible values of $R(na,t)$ are listed  
 in the fourth column of Table \ref{tableSchematic}. They are reproduced by the formula
\begin{equation}\label{RDef}
   R(na,t) =  \frac{r}{2}\left[1 - s(na,0)s(na,t)\right],\;\;\;\;\;\;\;\;\;\;(n\in \mathbb{Z}),
\end{equation}
 because the term  $1-s(na,0)s(na,t)$ takes the value $2$ if the opinion state $s(na,t)$ differs from its initial value (and the value zero otherwise).
  With these notations, the total flipping rate $w(na,t)$ of the binary variable $s(na,t)$ in the voter model under resetting is expressed  as
\begin{equation}\label{wDef}
\begin{split}
   w(na,t) :=& W(na,t) +R(na,t)\\
=& \frac{1}{2}\left[\tau^{-1}\left( 1- \frac{1}{2}s(x,t)\sum_{\epsilon = \pm 1}s((n+\epsilon) a,t)  \right) + r(1 - s(na,0)s(na,t))\right].
\end{split}
\end{equation}
 The possible values of $w(na,t)$ depend on the opinion states at time $t$ 
 at sites $(n-1)a, na, (n+1)a$, and on the initial opinion state at site $na$. The possible 
  values are listed and mapped to the associated configurations in Table \ref{tableSchematic}. To represent the configuration of the system graphically, we can
 symbolize opinion states by arrows (upward-pointing if the opinion state is positive and downward-pointing if the opinion state in negative). To calculate the flipping  rates, we need 
 to show at every vertex  a colored arrow symbolizing the current opinion state (say a red upward-pointing arrow if the opinion state is positive, and a blue downward-pointing arrow if the opinion state is negative), and a black arrow symbolizing the initial opinion state. The rates of the flipping processes of the opinion state at site $na$ listed in Table \ref{tableSchematic} are represented diagrammatically in Fig. \ref{figSchematic} \\

 The average opinion at time $t$ of the voter at site $x$ is the ensemble average of $s(x,t)$.  Let us denote it by
\begin{equation}\label{notS}
S(x,t) := \langle s(x,t) \rangle.
\end{equation} 
 This average opinion state is also called average magnetization (if the binary variables are interpreted as spins, the dynamics of the ordinary voter model is the zero-temperature Glauber dynamics).
 Another quantity of interest is  the spatial two-point function of the opinion state:
\begin{equation}\label{GDef}
G(x,y,t) := \langle s(x,t) s(y,t) \rangle,
\end{equation} 
which describes the extend to which two distant voters agree at time $t$.\\

 The configuration of the system at a given time can be described by a sequence of opinions as in Eq. (\ref{config}), or equivalently by the 
 list of pairs of nearest neighbors that have opposite opinions (together with the opinion of one voter, say at the origin). These pairs  are called the domain walls of the configuration.
 They separate domains on the lattice where the voters have a uniform opinion. 
 Consider the average density of domain walls at position 
 $x$ at time $t$, which we denote by $\rho(x,t)$. It is the probability that the spin variables at site $x$ and $x+1$ have opposite values
\begin{equation}\label{ProbComb}
 \rho(x,t)  = {\mathrm{Prob}}\left(\{s(x,t) = -1, s(x+a,t) = +1\}\right) + {\mathrm{Prob}}\left(\{s(x,t) = +1, s(x+a,t) = -1 \}\right).
\end{equation}
 In the particular case where the sites  $x$ and $y$ are nearest neighbors, the two-point function $G(x,y,t)$ allows to express the density of 
 domain walls. Indeed,
\begin{equation}\label{rhoDef}
\begin{split}
G(x,x+a,t) =& \langle s(x,t) s(x+a,t) \rangle \\
           =&   {\mathrm{Prob}}(s(x,t) = +1, s(x+a,t) = +1) + {\mathrm{Prob}}(s(x,t) = -1, s(x+a,t) = -1)\\
  &- {\mathrm{Prob}}(s(x,t) = -1, s(x+a,t) = +1) - {\mathrm{Prob}}(s(x,t) = -1, s(x+a,t) = +1) \\
   =& 1- 2\rho(x,t),\;\;\;\;\;\;\;\;\;\;\;\;\;\;\shoveright{x\in a \mathbb{Z}.}
\end{split}
\end{equation} 
We combined Eq. (\ref{ProbComb}) and  normalization of probability in the last step.\\

 We will use the following notations for the average opinion state  and two-point function  in the initial configuration:
 \begin{equation}\label{JDef}
S_0(x) := \langle s(x,0) \rangle,\;\;\;\;\;\;\;G_0(x,y) := \langle s(x,0)s(y,0) \rangle,\;\;\;\;\;\;\;x,y \in a\mathbb{Z}.
\end{equation} 
 The functions $S_0$ and $G_0$ are considered data of the problem: the initial configuration $\left\{s(x,0), x\in a\mathbb{Z} \right\}$ is random but drawn from a distribution that yields a fixed average opinion state and a fixed two-point function. 

\subsection{Average opinion state}\label{AM}

\subsubsection{Evolution equation}
When a binary opinion state $s(na,t)$ in $\{-1,+1\}$ changes, its value is shifted  by $-2s(na,t)$.  The evolution equation of the average opinion state follows as
\begin{equation}
\frac{\partial S}{\partial t} (na,t)= - 2 \langle s(na,t) w(na,t) \rangle.
\end{equation}
Substituting the flipping rate defined in Eq. (\ref{wDef})  and using the constraint $s(na,t)^2=1$ yields
\begin{equation}\label{evolMag}
\begin{split}
\frac{\partial S}{\partial t}(na,t) =& -  \left \langle\tau^{-1}\left( s(na,t)- \frac{1}{2}s(na,t)^2\sum_{j=\pm 1}s(na+ja,t)  \right) + r\left(s(na,t) - s(na,0)s(na,t)^2\right)  \right\rangle\\
=& \tau^{-1}\left[-  S(na,t) + \frac{1}{2}\left(S(na-a, t) + S(na+a, t)\right) \right]+r\left[ S_0(na)- S(na,t) \right],\;\;\;\;\; n\in {\mathbb{Z}}.
\end{split}
\end{equation}
%
 The  terms proportional to $\tau^{-1}$  on the r.h.s. correspond to the ordinary voter model, whose average opinion state satisfies a diffusion equation on the lattice.  The $r$-dependent terms are analogous  to a heat transfer 
  according to the Newton law of cooling:  
    the initial opinion-state profile $S_0$ is analogous to the temperature of a thermostat, and the resetting rate is analogous to the rate of thermal transfer between 
 the system and the thermostat.\\

Given a function  $f$ of the discrete space coordinate (and possibly other variables),  let us denote by $\hat{f}$ its Fourier transform:
\begin{equation}\label{FourierDef}
 \hat{f}(k):= \sum_{n\in \mathbb{Z}} f(na) e^{ikna},\;\;\;\;\;\;\;\;\;\; \shoveright{(k\in \mathbb{R}).}
\end{equation}
The Fourier transform is inverted by integrating over the first Brillouin zone $[-\pi/a, \pi/a]$:
\begin{equation}\label{FourierInv}
f(na) = \frac{a}{2\pi} \int_{-\frac{\pi}{a}}^{\frac{\pi}{a}}\hat{f}(k) e^{-ikna} dk,\;\;\;\;\;\;\;\;\;\;(n\in \mathbb{Z}).
\end{equation}
 With these notations, the Fourier transform of Eq. (\ref{evolMag}) reads
\begin{equation}\label{evolMagFourier}
\frac{\partial \hat{S}(k,t)}{\partial t} = \left[ \tau^{-1}\cos(ka) - (r + \tau^{-1}) \right]  \hat{S}(k,t)+r \hat{S_0}(k).
\end{equation}
 For a fixed momentum $k$, this ordinary differential equation is readily solved with the initial condition $\hat{S}(k,0) = \hat{S_0}(k)$:
\begin{equation}\label{Skt}
\begin{split}
 \hat{S}(k,t) = \exp &\left( [\tau^{-1}\cos(ka) - (r+\tau^{-1}) ]t \right)  \hat{S_0}(k) \\
&+r \int_{0}^t du\,  \hat{S_0}(k) \exp\left( [\tau^{-1}\cos(ka) - (r+\tau^{-1}) ] u\right).
\end{split}
\end{equation}
Fourier inversion yields the average opinion state at position $na$ and time $t$ as
\begin{equation}\label{explicitS}
\begin{split}
 S(na,t)   =&\frac{a}{2\pi}    \int_{-\frac{\pi}{a}}^{\frac{\pi}{a}} dk \exp\left( [\tau^{-1}\cos(ka) - (r+\tau^{-1}) ]t -ikna \right)  \hat{S_0}(k) \\
 &+ r \int_{0}^t du \frac{a}{2\pi}    \int_{-\frac{\pi}{a}}^{\frac{\pi}{a}}  dk   \hat{S_0}(k) 
                      \exp\left( [ \tau^{-1}\cos(ka) - (r+ \tau^{-1}) ] u -ik n a\right),\;\;\;\;\;\;    n\in {\mathbb{Z}}.
\end{split}
\end{equation}
 Let us inject the generating function of the modified Bessel functions of the first kind:
\begin{equation}\label{BesselGener}
e^{z\cos( ka)} = \sum_{ m \in \mathbb{Z}} I_m(z) e^{imka}.
\end{equation}
Permuting the summations and rearranging yields
\begin{equation}\label{convolExact}
\begin{split}
 S(na,t)   =& \frac{a}{2\pi}    \int_{-\frac{\pi}{a}}^{\frac{\pi}{a}} dk e^{  - (r+\tau^{-1}) t -ikna }\sum_{m\in \mathbb{Z}} I_m(\tau^{-1} t)   \hat{S_0}(k) e^{imka}\\
&+ r \int_{0}^t du \frac{a}{2\pi}    \int_{-\frac{\pi}{a}}^{\frac{\pi}{a}}  dk   \hat{S_0}(k) 
                       e^{(  - (r+ \tau^{-1})  u -ik n a)}  \sum_{m\in \mathbb{Z}} I_m(\tau^{-1} u)  e^{imka}\\
=& e^{  - (r+\tau^{-1}) t  }   \sum_{m\in \mathbb{Z}} I_m(\tau^{-1} t) \frac{a}{2\pi}    \int_{-\frac{\pi}{a}}^{\frac{\pi}{a}} dk
    \hat{S_0}(k) e^{i(m-n)ka}\\
&+ r \int_{0}^t du e^{ - (r+ \tau^{-1})  u}  \sum_{m\in \mathbb{Z}} I_m(\tau^{-1} u)   \frac{a}{2\pi}\int_{-\frac{\pi}{a}}^{\frac{\pi}{a}}  dk   \hat{S_0}(k)  e^{i(m-n)ka}\\
=& e^{  - (r+\tau^{-1}) t  }   \sum_{m\in \mathbb{Z}} I_m(\tau^{-1} t)  S_0( na-ma)\\
  & + r \int_{0}^t du e^{ - (r+ \tau^{-1})  u}  \sum_{m\in \mathbb{Z}} I_m(\tau^{-1} u) S_0(na-ma).\\
\end{split}
\end{equation}
 In the last step we used the identity $S_0(Na) = \frac{a}{2\pi}\int_{-\frac{\pi}{a}}^{\frac{\pi}{a}}  dk   \hat{S_0}(k)  e^{-iNka}$, which is the Fourier representation of the initial average opinion-state profile.
The average opinion-state profile at time $t$ is therefore the discrete convolution of $S_0$ 
  and the time-dependent kernel ${\mathcal{K}}_t$ defined as follows:
\begin{equation}\label{KDef}
\begin{split}
  S(na,t) &= \sum_{m\in \mathbb{Z}} {\mathcal{K}}_t(ma) S_0( na-ma),\;\;\;\;\;\;   \\
{\mathrm{with}}\;\;\;\;\;{\mathcal{K}}_t(na)&:= e^{  - (r+\tau^{-1}) t  }   I_n(\tau^{-1} t)
 +   r \int_{0}^t du e^{ - (r+ \tau^{-1})  u}  I_n(\tau^{-1} u),\;\;\;\;\;\;\;\;\;  \hspace{1cm}  (n\in {\mathbb{Z}}).\\
\end{split}
\end{equation}

Let us denote with a tilde the Laplace transform of any quantity depending on time (and possibly other variables):
\begin{equation}\label{LaplaceTransformDef}
\tilde{f}(s) := \int_0^\infty dt e^{-st} f(t).
\end{equation}
 The large-time limit ${\mathcal{K}}_\infty$ of the kernel ${\mathcal{K}}_t$ is expressed in terms of the  Laplace transforms
 of modified Bessel functions of the first kind. Using tabulated formulas listed in Appendix \ref{Laplace} (Eq. (\ref{Itilden})) yields:
\begin{equation}\label{Kinf}
\begin{split}
{\mathcal{K}}_\infty(na) =& r\tau \int_0^\infty dv e^{-(r\tau + 1)v} I_n(v)=  r\tau \tilde{I}_n( r\tau+ 1)\\
  =& \frac{r\tau}{\sqrt{(r\tau + 1)^2 - 1}}[(r\tau + 1) + \sqrt{(r\tau + 1)^2 - 1}]^{-|n|}.
\end{split}
\end{equation}

 The approach to the steady state of the kernel is described by the large-time behavior of the following quantity:
\begin{equation}\label{approach}
{\mathcal{K}}_t(na)   -  {\mathcal{K}}_\infty(na) = e^{  - (r+\tau^{-1}) t  }   I_n(\tau^{-1} t) - r \int_t^\infty du e^{ - (r+ \tau^{-1})  u}  I_n(\tau^{-1} u).
\end{equation}
 Let us take the Fourier representation:
 \begin{equation}
{\mathcal{K}}_t(na)   -  {\mathcal{K}}_\infty(na) = \frac{a}{2\pi}\int_{-\pi}^\pi  dk 
\left[ \hat{\mathcal{K}}_t(k)- \hat{\mathcal{K}}_\infty(k)\right] e^{-inka}.
\end{equation}
Using Eq. (\ref{BesselGener}) we can calculate  $\hat{\mathcal{K}}_t(k)$ using the generating 
 function of the modified Bessel functions:
\begin{equation}\label{KFour}
\begin{split}
\hat{\mathcal{K}}_t(k) =&  e^{  - (r+\tau^{-1}) t  }  \sum_{n\in\mathbb{Z}} I_n(\tau^{-1} t)e^{ikna}
 +   r \int_{0}^t du e^{ - (r+ \tau^{-1})  u}  \sum_{n\in\mathbb{Z}} I_n(\tau^{-1} u)e^{ikna}\\
=& e^{  -[ r +\tau^{-1}(1-\cos(ka) )] t  } 
 +   r \int_{0}^t du e^{  -[ r +\tau^{-1}(1-\cos(ka) )] u  }\\
=& \frac{r + \tau^{-1}( 1- \cos(ka)) e^{-[ r +\tau^{-1}(1-\cos(ka) )] t}}{r +\tau^{-1}(1-\cos(ka) )}.\\
\end{split}
\end{equation}
 The large-time limit of the Fourier transform  of the kernel follows as
\begin{equation}\label{KFourInf}
\begin{split}
\hat{\mathcal{K}}_\infty(k) =& \frac{r}{r +\tau^{-1}(1-\cos(ka) )}.
\end{split}
\end{equation}
Hence
\begin{equation}\label{FourierRepd1}
\begin{split}
\left|{\mathcal{K}}_t(na)   -  {\mathcal{K}}_\infty(na) \right|=& \left|\frac{a}{2\pi}\intB dk\left[ \left( \hat{\mathcal{K}}_t(k)- \hat{\mathcal{K}}_\infty(k)\right)\right]e^{-inka}\right|\\
=&\left|\frac{a}{2\pi}\intB dk \left[ 
\frac{\tau^{-1}( 1- \cos(ka)) e^{-[ r +\tau^{-1}(1-\cos(ka) )] t}}{r +\tau^{-1}(1-\cos(ka) )}
\right]e^{-inka}\right|\\
=& e^{-rt}\left|\frac{a}{2\pi}\intB dk 
\frac{\tau^{-1}( 1- \cos(ka)) e^{-\tau^{-1}(1-\cos(ka) ) t}}{r +\tau^{-1}(1-\cos(ka) )}
e^{-inka}\right|\\
\leq & e^{-rt}\int_{-\pi}^\pi \frac{dq}{2\pi} \left|e^{-\tau^{-1}(1-\cos(q) ) t}
\frac{\tau^{-1}( 1- \cos(q))}{r +\tau^{-1}(1-\cos(q) )}\right|\\
\leq &e^{-rt} \int_{-\pi}^\pi  \frac{dq}{2\pi}  e^{-\tau^{-1}(1-\cos(q) ) t}\\
= & e^{-(r+\tau^{-1})t} I_0( \tau^{-1}t).
\end{split}
\end{equation}
Using the equivalent
\begin{equation}\label{largeArg}
 I_n(z) \underset{z\to \infty}{\sim} \frac{e^z}{\sqrt{2\pi z}},\;\;\;\;\;\;\;\;\hspace{1cm}\shoveright{(n\in\mathbb{Z}),}
\end{equation}
we notice that the upper bound is equivalent to $(2\pi \tau^{-1} t)^{-\frac{1}{2}}e^{-rt}$
 when time $t$ is large compared to the time scale $\tau$.\\

 Moreover,  we can obtain an upper bound on 
 $|S(na,t)-S(na,\infty)|$  by using the Fourier transform of Eq. (\ref{KDef}). 
 As the average opinion state profile is a convolution, its Fourier 
 transform at time $t$ is given by the product of the Fourier transforms of the kernel 
$\mathcal{K}_t$ and the initial average opinion state:  
\begin{equation}\label{SFour}
\begin{split} 
\hat{S}(k,t) =& \hat{S}_0(k)\hat{\mathcal{K}}_t(k),\;\;\;\;\;\;\hspace{1cm}(t\geq 0). 
\end{split}
\end{equation}

Combining Eqs (\ref{KFour},\ref{KFourInf},\ref{SFour},) and using the Fourier representation of the kernel as in Eq. (\ref{FourierRepd1}) 
 yields
\begin{equation}\label{upperBound1d}
\begin{split}
\left|S(na,t) - S(na,\infty) \right|=& \left|\frac{a}{2\pi}\intB dk\left[ \hat{S}_0(k)\left( \hat{\mathcal{K}}_t(k)- \hat{\mathcal{K}}_\infty(k)\right)\right]e^{-inka}\right|\\
=&\left|\frac{a}{2\pi}\intB dk \left[ \hat{S}_0(k)
\frac{\tau^{-1}( 1- \cos(ka)) e^{-[ r +\tau^{-1}(1-\cos(ka) )] t}}{r +\tau^{-1}(1-\cos(ka) )}
\right]e^{-inka}\right|\\
=& e^{-rt}\left|\frac{a}{2\pi}\intB dk \hat{S}_0(k)
\frac{\tau^{-1}( 1- \cos(ka)) e^{-\tau^{-1}(1-\cos(ka) ) t}}{r +\tau^{-1}(1-\cos(ka) )}
e^{-inka}\right|\\
\leq & e^{-rt}\frac{a}{2\pi}\intB dk \left|\hat{S}_0(k)e^{-\tau^{-1}(1-\cos(ka) ) t}
\frac{\tau^{-1}( 1- \cos(ka))}{r +\tau^{-1}(1-\cos(ka) )}\right|\\
\leq &e^{-rt}\frac{a}{2\pi}\intB dk \left|\hat{S}_0(k)\right|e^{-\tau^{-1}(1-\cos(ka) ) t}.
\end{split}
\end{equation}
 Moreover, the upper bound is uniform in the integer $n$ labeling the site. It goes to zero if $\hat{S}_0$ is integrable on $[-\pi/a,\pi/a]$.\\

\subsubsection{Example: a single  decided voter at the origin}\label{exampleS}
 Consider an initial condition with independent undecided voters, except for a single decided voter with positive opinion state at the origin. If $n\neq 0$, the opinion state $s(na,0)$ is a centered Bernoulli random variable. On the other hand $s(0,0)=1$. Hence  
\begin{equation}
S_0(ma) := \delta_{m,0},\;\;\;\;\;\;\;\;\;\;(m  \in {\mathbb{Z}}).
\end{equation}

\begin{figure}
\begin{center}
\includegraphics[width=16cm]{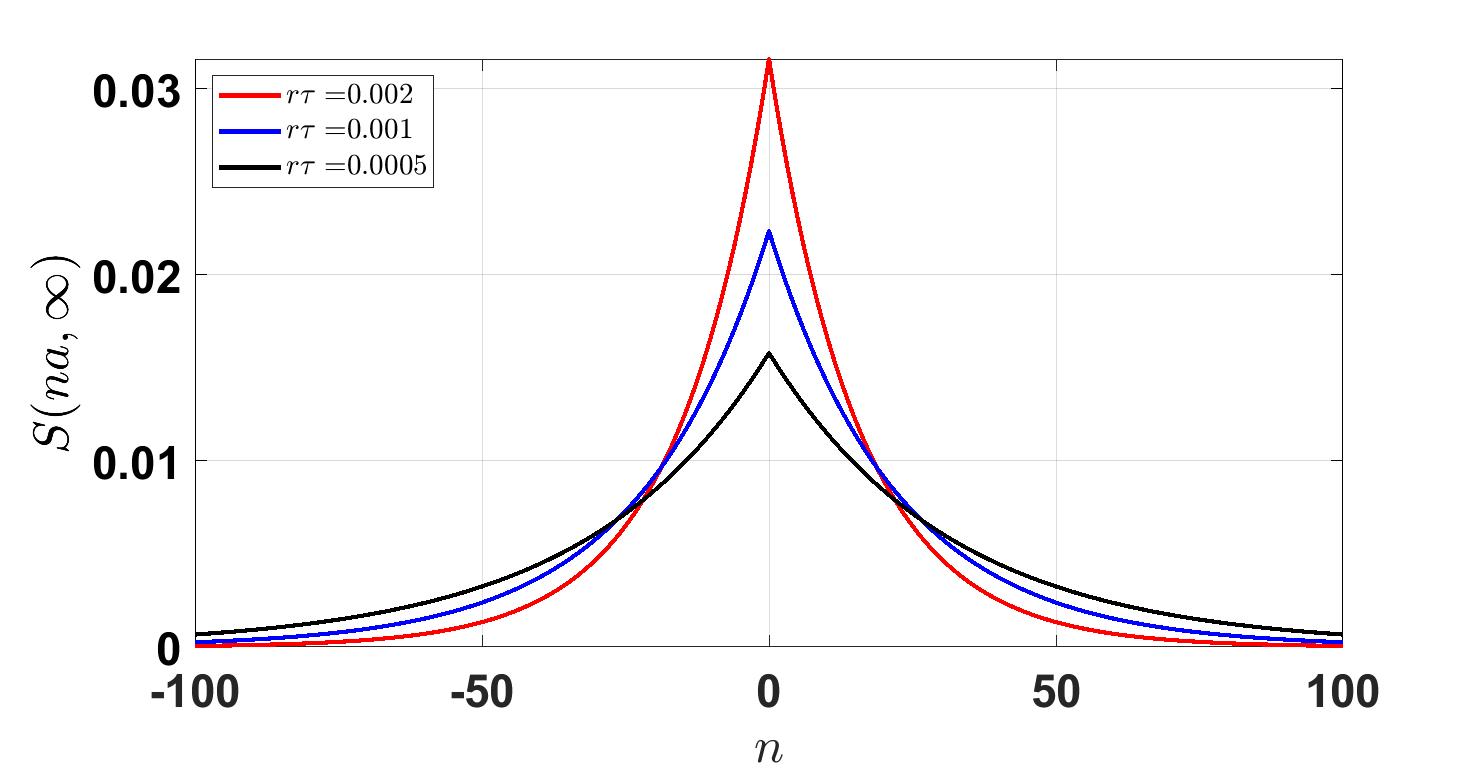} 
\caption{{\bf{The average opinion-state profile in the steady state for initial conditions consisting of a single decided voter with positive opinion state at the origin (the other voters being undecided).}}}
 \label{figMagnetizationSingleDecided}
\end{center}
\end{figure}

In this case the  average opinion state at site labeled $n$ coincides with  the quantity ${\mathcal{K}}_t(na)$ defined in Eq. (\ref{KDef}):
\begin{equation}\label{singleDecided}
 S(na,t)  =
 e^{  - (r+\tau^{-1}) t  }   I_n(\tau^{-1} t)  
   + r \int_{0}^t du e^{ - (r+ \tau^{-1})  u}  I_n(\tau^{-1} u),\;\;\;\;(n\in{\mathbb{Z}}).
\end{equation}
  The r.h.s. does not depend on the lattice spacing $a$, which is expected because the parameter $a$ does not appear in the definition of the flipping rates.  
 The steady state of the average opinion state follows from Eq. (\ref{Kinf}) as
  an exponentially-decaying function of the distance to the origin (plotted in Fig. \ref{figMagnetizationSingleDecided}):
 \begin{equation}\label{lambdaDef}
\begin{split}
S(na,\infty) &= \mathcal{K}_t(na) = r\tau \tilde{I}_n(r\tau + 1)= \frac{r\tau}{\sqrt{(r\tau + 1)^2 - 1}} \lambda(r,\tau)^{-|n|} \\ 
{\mathrm{with}}\;\;\;\;\;\lambda(r,\tau)&:= r\tau + 1 + \sqrt{ (r\tau + 1)^2 -1}.
\end{split}
 \end{equation}
  In dimension one, the above result solves Problem (I$_{(r)}$) defined in the introduction.  
 Indeed, the average opinion  state of the initially decided voter does not relax to zero in the steady state as it does in the ordinary model. In the limit of low resetting rate, it is proportional to the square root of the resetting rate:
 \begin{equation}\label{equiv1dS}
S(0,\infty) \underset{r\ll \tau^{-1}}{\sim} \sqrt{\frac{r\tau}{2}}.
 \end{equation}

%

In the initial state, the average opinion-state profile is nonnegative and normalized, indeed $\sum_{n\in\mathbb{Z}}S(na,0)=1$. 
 Moreover, Eq. (\ref{evolMag}) becomes in this case
\begin{equation}\label{evolMagSingleDecided}
\begin{split}
\frac{\partial S}{\partial t}(na,t)
=& \tau^{-1}\left[-  S(na,t) + \frac{1}{2}\left(S(na-a, t) + S(na+a, t)\right) \right]- r S(na,t)+ r\delta_{n0},\;\;\;\;\;\;\;\;\;\ (n\in {\mathbb{Z}}).
\end{split}
\end{equation}
 The terms carrying  a factor of $\tau^{-1}$ describe the motion of a diffusive random walker on the lattice $a\mathbb{Z}$.  The $r$-dependent terms $r( \delta_{n,0} - S(na, t))$ in Eq. (\ref{evolMag}) correspond to resetting the position of the random walker to the origin between times $t$ and $t+dt$ with probability $rdt$. 
The quantity $S(na,t)={\mathcal{K}}_t(na) $ is the probability of presence at site $na$ and time $t$ of a diffusive random walker on the lattice, starting at the origin at time $0$ and stochastically reset to its initial position at rate $r$.
The average opinion-state profile is therefore normalized at all times:
\begin{equation}
  \sum_{n\in\mathbb{Z}}S(na,t)= \sum_{n\in\mathbb{Z}} {\mathcal{K}}_t(na) = 1.\;\;\;\;\;\;\;\;\hspace{1cm}(t\geq 0).
\end{equation}

{\bf{Continuum limit.}}  Let us take the
 continuum limit as follows: the lattice
 spacing $a$ and the characteristic time $\tau$ both go to zero, while the resetting rate $r$ and the diffusion constant
 \begin{equation}
  D:= \frac{ a^2}{2\tau }
 \end{equation}
 are kept fixed.\\

  Let us consider a large position $x$ (on the scale of the lattice spacing $a$):
\begin{equation}
 x := na,\;\;\;\;\;\; |n|\gg 1.
\end{equation}
 Expanding the expression of the  quantity $\lambda(r,\tau)$ given in Eq. (\ref{lambdaDef}) in powers of $r\tau$ yields
  \begin{equation}
\lambda(r,\tau) =1 + \sqrt{ 2r\tau} +  o(\sqrt{r\tau}).
 \end{equation}
In the large-distance regime, the spatial decay of the average opinion state is obtained as
\begin{equation}
\lambda(r,\tau)^{-|n|} = \exp\left( - |n| \log[\lambda(r,\tau) ] \right) 
\sim \exp\left(-\frac{|x|}{a}\sqrt{2r\tau}\right) = e^{-|x|\sqrt{\frac{r}{D}}},
\end{equation} 
 which is the simplest exponential decay that can be expressed using the position, resetting rate and diffusion coefficient,  
 for dimensional reasons. Introducing a factor of $a^{-1}$, we obtain the diffusive limit of the average opinion state
 per unit of length as 
\begin{equation}\label{continuum1d}
a^{-1} S(na, \infty) =  \frac{r\tau}{\sqrt{a^2[(r\tau + 1)^2 - 1]}} \lambda(r,\tau)^{-|n|}
\underset{a,\tau \to 0, x=na, D = \frac{a^2}{2\tau}}{\sim} \frac{1}{2}\sqrt{\frac{r}{D}} e^{-x\sqrt{\frac{r}{D}}}.
\end{equation}
As expected, the diffusive limit of the random walk on the lattice with resetting reproduces  the  steady-state  probability density of a diffusive random random walker 
  on a line under stochastic resetting to the origin (\cite{evans2011diffusion,evans2011optimal}, for a review see Section 2.3 of \cite{topical}).

\subsubsection{Example with infinite total opinion state: decided voters on a half line}

\begin{figure}
\begin{center}
\includegraphics[width=16cm]{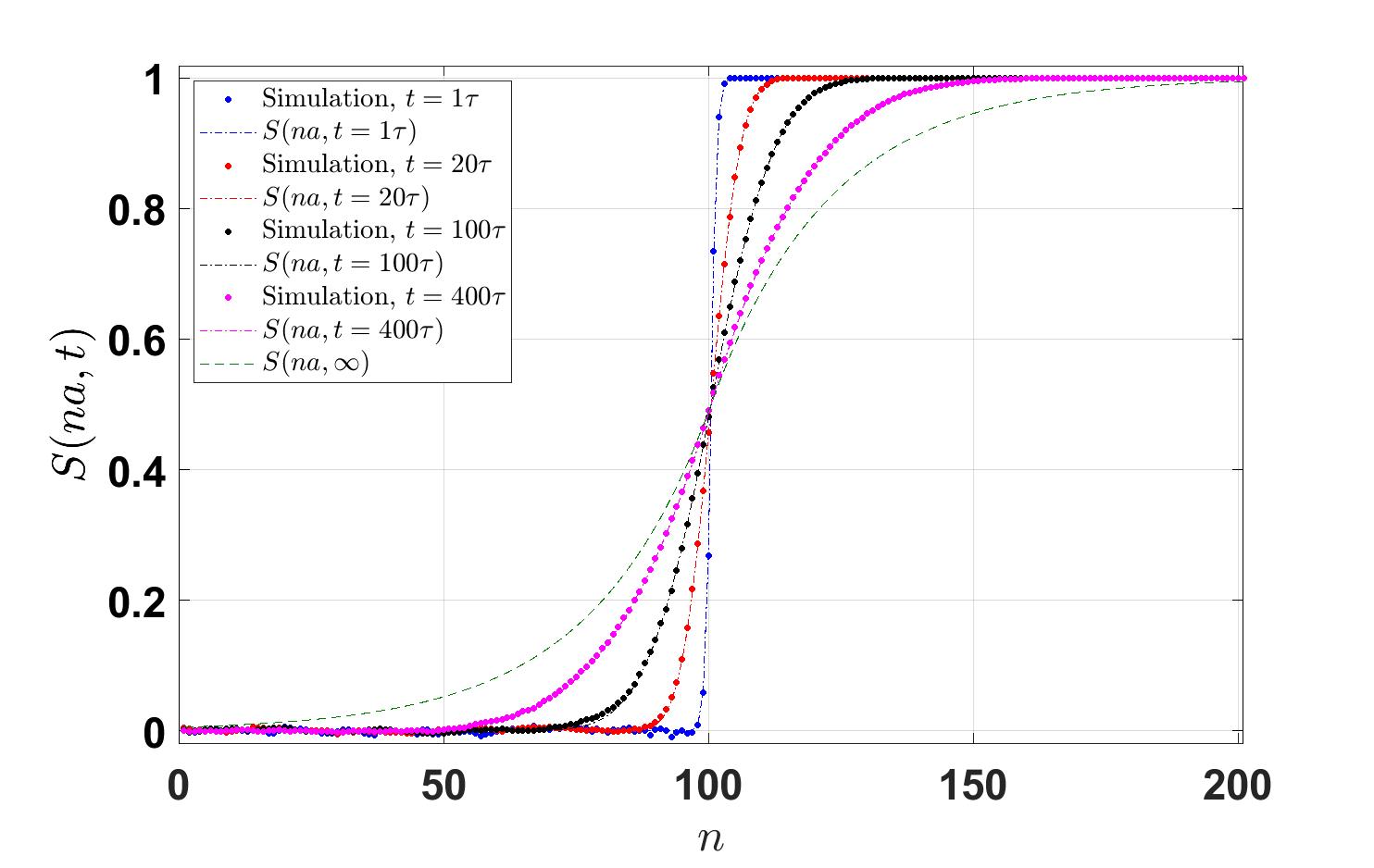} 
\caption{{\bf{Average opinion state for initial conditions  with undecided voters in the negative half line and decided  voters in the positive half line.}}   The parameters are  the resetting rate $r=10^{-3}$ and the time $\tau=1$.  Dots correspond to a direct simulation of the model (with $-200\leq n \leq 200$), averaged over  $N=100,000$ samples of the initial configuration with independent undecided voters in the negative half-line. The predicted value $S(na,t)$ is based on  Eq. (\ref{relaxationStepEq}).}
 \label{relaxationStep}
\end{center}
\end{figure}

  Not all the possible initial configurations of the voter model satisfy 
$\sum_{n\in\mathbb{Z}}|S_0(na)|<\infty$. As an example, consider an initial condition with independent undecided voters at sites with negative indices, and
 decided voters with positive opinion state occupying at sites with nonnegative indices:
 \begin{equation}
 S_0(m) := \mathbf{1}( m \geq 0 ),\;\;\;\;\;\;\;\;\;\;\hspace{1cm}(m\in\mathbb{Z}).
\end{equation}
 This configuration has infinite total opinion state, but it is still amenable to explicit calculation. From the discrete convolution in Eq. (\ref{convolExact}) we obtain
\begin{equation}
S(na,t) = e^{  - (r+\tau^{-1}) t  }   \sum_{m = -\infty}^n I_m(\tau^{-1} t) \\
   + r \int_{0}^t du e^{ - (r+ \tau^{-1})  u}  \sum_{m = -\infty}^n I_m(\tau^{-1} u).
\end{equation}
Using the identities
\begin{equation}
   \sum_{m=1}^\infty I_m(z) = \frac{1}{2}\left(  e^z - I_0(z) \right), \;\;\;\;\; I_{n} = I_{-n},\;\;\;\;\;\;(n \in\mathbb{Z}),
\end{equation}
 we obtain the expression of the average opinion state at site $na$ as  a sum of a finite number of terms:

\begin{equation}\label{relaxationStepEq}
\begin{split}
S(0,t)=&  \frac{1}{2} +  \frac{1 }{2}e^{ - (r+ \tau^{-1})  t} I_0(\tau^{-1} t)  
    +  \frac{r}{2}  \int_{0}^t du e^{ - (r+ \tau^{-1})  u} I_0(\tau^{-1} u),\\
S(na,t) =& S(0, t)  +   e^{ - (r+ \tau^{-1})  t}\sum_{m=1}^{|n| }   I_m(  \tau^{-1} t)  + r \int_0^t du  e^{ - (r+ \tau^{-1})  u}\sum_{m=1}^{|n| }   I_m(  \tau^{-1} u), \;\;\;\;\;\;(n>0),\\
S(na,t)
 =&  S(0, t)  -   e^{ - (r+ \tau^{-1})  t}\sum_{m=0}^{|n|-1 }   I_m(  \tau^{-1} t)  - r \int_0^t du  e^{ - (r+ \tau^{-1})  u}\sum_{m=0}^{|n|-1 }   I_m(  \tau^{-1} u), \;\;\;\;\;\;(n<0).\\ 
\end{split}
\end{equation}
 The average opinion state is plotted as a function of $n$ in Fig. \ref{relaxationStep}, together with   the result of direct numerical simulations performed for a system with a finite number of sites\footnote{The algorithm is as follows. The system consists of $2K+1$ voters.   The configuration is  initialized: $s(na,0)=1$ for $0\leq n \leq K$, and 
 the  value of $s(na,0)$ is drawn from  a centered Bernoulli distribution for $ -K\leq n\leq -1$. Time is initialized as $t=0$. The flipping rates and their sum $\Phi:=\sum_{n=-K}^K w(na,t)$ are computed. The time $t_1$ of the first flipping event is drawn from an exponential distribution with parameter $\Phi$.  The voter whose opinion state flips at $t_1$ is then drawn from a discrete 
 distribution (the voter at position $n$ changes its opinion state with probability $w(na,t)/\Phi$).
 The opinion state of this voter is flipped, which yields the configuration of the system at time $t_1$. Time is changed to $t=t_1$, the new flipping rates are calculated, and the process is iterated. This algorithm is run for $N$ independent samples of the system and the estimated  average opinion state $S(na,t)$ is the average of $s(na,t)$ over the samples.}.\\
 
Let us take the large-time limit.  The equivalents of the Bessel functions Eq. (\ref{largeArg})
 imply that that the terms in Eq. (\ref{relaxationStepEq}) that are not in integral form go to zero. The steady state is again expressed in terms of Laplace transform of Bessel functions:
\begin{equation}
\begin{split}
S(0,\infty) &= \frac{1}{2} +\frac{r\tau}{2}\int_0^\infty dv e^{-(r\tau +1)v} I_0(v) = \frac{1}{2} + \frac{r\tau}{2\sqrt{(r\tau+1)^2 -1}},\\
S(na,\infty) &=S(0,\infty) +  r\tau  \sum_{m=1}^{n} \tilde{I}_m(r\tau + 1)\\
  &= S(0,\infty) + \frac{r\tau}{ \sqrt{(r\tau+1)^2 -1}} \frac{ 1- ( r\tau + 1+ \sqrt{(r\tau+1)^2 -1} )^{-n}}{ r\tau + \sqrt{(r\tau+1)^2 -1}},\;\;\;\;\;(n>0),\\
S(na,\infty) &=S(0,\infty) -  r\tau \sum_{m=0}^{n-1} \tilde{I}_m(r\tau + 1)\\
  &= S(0,\infty) - \frac{r\tau( r\tau+ 1 +\sqrt{(r\tau+1)^2 -1} )}{ \sqrt{(r\tau+1)^2 -1}} \frac{ 1- ( r\tau + 1+ \sqrt{(r\tau+1)^2 -1} )^{-|n|}}{ r\tau + \sqrt{(r\tau+1)^2 -1}},\;\;\;\;\;(n<0).
\end{split}    
\end{equation}
 This steady state is shown as a green dashed line on Fig. \ref{relaxationStep}.

\subsection{Two-point function and density of domain walls}\label{twoPoint}

\subsubsection{Evolution equation of the two-point function}

 Consider the two-point function $G(x,y,t)=\langle s(x,t) s(y,t)\rangle$.
 If $x$ and $y$ are equal, $G$ is constant because of the constraint $s(x,t)^2=1$.   On the other hand, if $x=na$ and $y=ma$ are two distinct sites on the lattice, 
the value of $G(x,y,t)$ changes when any of the voters at position $x$ or $y$ changes its opinion state:
\begin{equation}\label{evolG}
\frac{\partial G}{\partial t}(na,ma,t) =
- 2 \langle s(na,t) s(ma,t)[ w(na,t)  + w(ma,t)] \rangle, \;\;\;\;\;\;\;(n,m\in\mathbb{Z},\, n\neq m).
\end{equation}
Substituting the expression of the flipping rate given in Eq. (\ref{wDef}) and using the constraint  $s(na,t)^2 = 1$ we obtain   
\begin{equation}\label{Gxy}
\begin{split}
2 s(na,t) s(ma,t)  w(na,t) =& \tau^{-1} s(na,t) s(ma,t) - \frac{\tau^{-1}}{2} s(ma,t) \sum_{z= na\pm a} s(z,t)\\
 &+ r s(na,t) s(ma,t) - r s(ma,t) s(na,0).
\end{split}
\end{equation}
Taking the ensemble averages of both sides, we notice that one term proportional to the resetting rate
 involves the following  two-time two-point function:\\
\begin{equation}\label{defH}
 H(na,ma,t) := \langle s( na,t) s(ma,0) \rangle.
\end{equation}
 We will use this notation for any integer values of $m$ and $n$, but the terms that appear in the evolution equation of the two-point function $G$ correspond to $n\neq m$.\\  

With this notation, Eq. (\ref{Gxy}) implies
\begin{equation}\label{GxyBis}
\begin{split}
2 \langle s(na,t) s(ma,t)  w(na,t)\rangle =& \tau^{-1} G(na,ma,t) - \frac{\tau^{-1}}{2} \sum_{z= na\pm a} G(z, ma,t)\\ 
&+ r G(na,ma,t) - r H(ma,na,t),\;\;\;\;\;\;\;\;\hspace{1cm}(n,m\in\mathbb{Z}, \;n \neq m).
\end{split}
\end{equation} 

The quantity  $2 \langle s(x,t) s(y,t) w(y,t) \rangle$ is obtained by permuting $n$ and $m$ in the above expression, 
 which yields the following evolution equation:
\begin{equation}\label{evolGH}
\begin{split}
  \frac{\partial G}{\partial t}(na,ma,t) =& -2( r +\tauInv) G( na, ma, t)\\
 &+ \frac{\tauInv}{2}\left[ G( (n-1)a, ma,t) +  G( (n+1)a, ma,t) +  G( na, (m-1)a,t) +  G( a, (m+1)a,t) \right]\\
&+r [H( na, ma, t )+ H( ma, na, t )],
 \;\;\;\;\;\;\;\;\;\;\;\;\;\;\hspace{1cm}(m,n\in \mathbb{Z},\;n\neq m).
\end{split}
\end{equation}
On the other hand, the two-point function $G$ at coincident points is constant:
\begin{equation}\label{constraint}
G(na, na,t) = \langle  s( na,t)^2\rangle=1,\;\;\;\;\;\;\;\;\;\;\hspace{1cm} (n \in \mathbb{Z}).
\end{equation}
To solve the evolution equation of $G$,  we will  need to work out the evolution equation of the two-time  two-point function $H$ defined in Eq. (\ref{defH}). The evolution is induced by the flipping rate of
 the opinion state at site $na$:
\begin{equation}
           \frac{\partial H}{\partial t}(na,ma,t) = -2  \langle w(na,t) s(na,t) s(ma,0)   \rangle,\;\;\;\;\;\;\;\;\;(m,n\in\mathbb{Z}) .
\end{equation}
 This equation holds even for $n=m$, because the quantity $s(y,0) = S_0(y)$ does not evolve in time. 
  Substituting the expression of the flipping rate given in Eq. (\ref{wDef}) and using again the constraint  $s(na,t)^2 = 1$ yields
\begin{equation}\label{evolH}
\begin{split}
\frac{\partial H}{\partial t} (na,ma,t)=&
- \langle  s(na,t) \left[   \tau^{-1}\left( 1- \frac{1}{2}s(na,t)\sum_{z=na\pm a}s(z,t)  \right) + r(1 - s(na,0)s(na,t))\right] s(ma,0)   \rangle\\
=& -   \tau^{-1}\langle s(na,t) s(ma,0)  \rangle +\frac{\tau^{-1}}{2}\sum_{z=na\pm a}\langle s(z,t) s(ma,0)   \rangle \\
 &- r\langle s(na,t) s(ma,0)  \rangle 
 +     r\langle s(na,0) s(ma,0)  \rangle\\
=&- (r+\tauInv) H(na,ma,t) + \frac{\tauInv}{2} \sum_{z=x\pm a }H(na,ma,t)  
 +     r \delta_{mn},\\
{\mathrm{with}} &\;\;\;H(na,ma,0) = \langle s(na,0) s(ma,0)  \rangle = \delta_{nm},\;\;\;\;\;\;\;(n,m\in\mathbb{Z}).
\end{split}
\end{equation}
For any fixed $m$, we notice that $H(na,ma,t)$ satisfies the evolution equation of a diffusive random walker on the lattice, stochastically reset to its initial position $ma$.
At this point, let us specialize to the case of initially-undecided voters.

\subsubsection{Solution for initially undecided voters}

 Let us assume that the voters are independent and  undecided in the initial state: the initial 
 opinion states of the voters are independent centered Bernoulli random variables. Hence
\begin{equation}
\begin{split}
S_0(na) &= 0,\\
G_0(na,ma) &= \langle s(na,0) s(ma, 0)\rangle = \delta_{nm}+ (1-\delta_{nm})S_0(na)S_0(ma) = \delta_{nm},\;\;\;\;\;\;\;\;(m,n\in\mathbb{Z}).
\end{split}
\end{equation}
 
 This choice of initial conditions implies that  the system is translationally invariant.
  The value of the two-point function depends only on the separation of the voters (and on time). Hence there exist functions  $\mathcal{H}$ and $\mathcal{G}$ of two variables  such that:
\begin{equation}
\begin{split}
 H(na,ma, t) &= \mathcal{H}( (m-n) a, t ),\\ 
 G(na,ma,t) &= \mathcal{G}( (m-n) a, t ), \;\;\;\;\;\;\;\;\;\;\hspace{1cm}(m,n\in {\mathbb{Z}}, t\geq 0).
\end{split}
\end{equation}
 The initial condition $G_0(na,ma) = \delta_{nm}$ can be rewritten as
\begin{equation}
\mathcal{G}( na,0 ) = \delta_{n,0}, \;\;\;\;\;\;\;\;\;\;\;\;(n\in \mathbb{Z}).
\end{equation}

 The evolution equation Eq. (\ref{evolGH}) derived above holds for two distinct points, $n\neq m$, hence 
\begin{equation}\label{evolGcal1d}
\begin{split}
\frac{\partial \mathcal{G}}{\partial t}( na, t) =& -2( r + \tauInv) \mathcal{G}( na, t) 
+ \tau^{-1}[ {\mathcal{G}}( (n-1)a, t) + {\mathcal{G}}( (n+1)a,t) ] \\
 &+ 
   r\mathcal{H}( na, t) + r  \mathcal{H}( -na, t ), \;\;\;\;\;\;\;\;\;\hspace{1cm}(n\in \mathbb{Z}, \;n\neq 0). 
\end{split}
\end{equation}
  The constraint expressed in Eq. (\ref{constraint}) reads 
\begin{equation}
\mathcal{G}(0,t) =1,\;\;\;\;\;\;\;\;\;\;(t\geq 0).
\end{equation}
 To ensure this condition holds throughout the evolution of the system, let us add an (unknown) source term $J(t)$  at the origin:
\begin{equation}\label{evolGsource}
\begin{split}
\frac{\partial \mathcal{G}}{\partial t}( na, t) =& -2( r + \tauInv) \mathcal{G}( na, t) +
 \tau^{-1}[ {\mathcal{G}}( (n-1)a, t) + {\mathcal{G}}( (n+1)a,t) ] \\
  &+ r\mathcal{H}( na, t) + r  \mathcal{H}( -na, t )+ J(t) \delta_{n,0},\;\;\;\;\;\;(n\in \mathbb{Z}). 
\end{split}
\end{equation}
  This technique was introduced in \cite{frachebourg1996exact} to solve the  voter model  without resetting (see Chapter 8.2 in \cite{kineticView} for a review), which is recovered by setting the resetting rate $r$ to zero in the above equation.\\

 As Eq. (\ref{evolGsource}) holds for all values of the discrete space coordinate, we can take its Fourier transform. As we have noticed from Eq. (\ref{evolH}), the quantity  $H(na,ma,t)$ is the probability of presence at $na$ of a random walker stochastically reset to its initial position $ma$. It satisfies the parity property $H( (m-n)a,ma,t) = H( (m+n)a,ma,t)$ for every integer $n$, which implies $\mathcal{H}( na, t) = \mathcal{H}( -na, t )$. The Fourier transform of Eq. (\ref{evolGsource}) therefore reads
 \begin{equation}\label{evolGcal}
\begin{split}
\frac{\partial \hat{\mathcal{G}}}{\partial t}( k,t) &= -2(r +\tauInv) \hat{\mathcal{G}}( k,t) +
 2 \tauInv \cos( ka )  \hat{\mathcal{G}}(k,t) + 2r \hat{\mathcal{H}}(k,t) + J(t),\\
{\mathrm{with}}\;\;\;\;\hat{\mathcal{G}}( k,0) &=1,\;\;\;\;\;\;\;\;\;\hspace{1cm}k\in \left[-\frac{\pi}{a},\frac{\pi}{a}   \right].
\end{split}
\end{equation}
  Solving this ordinary differential equation yields the expression of the two-point function in Fourier space at time $t$ in terms of the unknown current $J$ (and the unknown function $\mathcal{H}$):
\begin{equation}\label{Gkt}
\begin{split}
\hat{\mathcal{G}}( k,t) =& e^{2 \alpha(k) t} +2r \int_0^t du e^{2\alpha(k)(t-u)} \hat{\mathcal{H}}(k,u)
 + \int_0^t du  e^{2\alpha(k)(t-u)} J(u), \\
{\mathrm{where}}&\;\;\;\;\;\;\alpha(k) = \tau^{-1}\cos(ka )    - (r+\tau^{-1}).
\end{split}
\end{equation}

At this point we need the Fourier transform of the function $\mathcal{H}$. 
  The evolution equation of the two-time two-point function $H$ (Eq. (\ref{evolH})) is readily rewritten in terms of the function $\mathcal{H}$ as
\begin{equation}\label{evolHcal1d}
\begin{split}
\frac{\partial \mathcal{H}}{\partial t}(na,t) &=
- (r+\tauInv) \mathcal{H}(na, t)+ \frac{\tauInv}{2} \left[\mathcal{H}((n+1)a, t) +\mathcal{H}((n-1)a, t) \right]
 +     r \delta_{n0},\\
{\mathrm{with}} &\;\;\;\mathcal{H}( na, 0)= \delta_{n0},\;\;\;\;\;\;\;\;\;\;\;\;\hspace{1cm}(n\in\mathbb{Z}).
\end{split}
\end{equation}
 This equation is identical to  the one satisfied by the average opinion state (Eq. \ref{evolMag}) with a single decided voter at the origin in the initial state. The Fourier transform with the initial condition $\hat{\mathcal{H}}(k,0) = 1$ is therefore obtained directly from Eq. (\ref{Skt}) as
\begin{equation}\label{Hkt}
\hat{\mathcal{H}}(k,t) =  \exp \left( [\tau^{-1}\cos(ka) - (r+\tau^{-1}) ]t \right) \\
+r \int_{0}^t du\,  \exp\left( [\tau^{-1}\cos(ka) - (r+\tau^{-1}) ] u\right).
\end{equation}

 Coming back  to Eq. (\ref{Gkt}) and inverting the Fourier transform yields
\begin{equation}\label{solGcalReal}
\begin{split}
{\mathcal{G}} ( na,t) =& e^{-2(r+\tauInv ) t}I_n( 2\tauInv t) +2r \frac{a}{2\pi} \intB dk \int_0^t du e^{2\alpha(k)(t-u) -inka} \hat{\mathcal{H}}(k,u)\\
 &+ \int_0^t du  e^{-2(r+\tauInv ) (t-u)}I_n( 2\tauInv (t - u))  J(u),\\
\end{split}
\end{equation}
 where in the first and last term we have used the identity $e^{2\tau^{-1}(t-u)\cos(ka)} = \sum_{n\in\mathbb{Z}} I_n( 2\tau^{-1}(t-u)) e^{inka}$, which is the generating function of the modified Bessel functions of the first kind defined in Eq. (\ref{BesselGener}).
We have to adjust the current $J$ so that the condition  ${\mathcal{G}} ( 0,t) =1$ is satisfied at all times:
\begin{equation}
\begin{split}
1 =& e^{-2(r+\tauInv ) t}I_0( 2\tauInv t) +2r \frac{a}{2\pi} \intB dk \int_0^t du e^{2\alpha(k)(t-u)} \hat{\mathcal{H}}(k,u)\\
 &+ \int_0^t du  e^{-2(r+\tauInv ) (t-u)}I_0( 2\tauInv (t - u))  J(u).\\
\end{split}
\end{equation}
Let us take the Laplace transform, which maps the convolution product to an ordinary product, leading to
\begin{equation}\label{exprJs}
\begin{split}
\frac{1}{s}=&  \tilde{\varphi}_0(s) +2r \Psi_0(s) + \tilde{\varphi}_0(s)  \tilde{J}(s),\\
\end{split}
\end{equation}
with the notations
\begin{equation}\label{varPhiDef}
\begin{split}
&\;\;\;\;\; \varphi_n(t):=  e^{-2(r+\tauInv ) t }I_n( 2\tauInv t ),\\
 {\mathrm{and}}&\;\;\;\;\;\Psi_n( s):= \frac{a}{2\pi} \intB dk e^{-inka} \frac{1}{s - 2\alpha(k)}  \tilde{\hat{\mathcal{H}}}(k,s),\;\;\;\;\;\;(n\in\mathbb{Z}).\\
\end{split}
\end{equation}
 Moreover, the expression of the two-point function in Eq. (\ref{solGcalReal}) can be used to relate the Laplace transform of the current $J$ 
 to the one of the density of domain walls.

\subsubsection{The steady-state density of domain walls}

In a translationally-invariant system, the density of domain walls is a function of time only, call it $\varrho$:
\begin{equation}
\begin{split}
\rho( na, t ) &= \varrho(t),\;\;\;\;\;(n\in {\mathbb{Z}}). 
\end{split}
\end{equation}

\begin{figure}[H]
\begin{center}
\includegraphics[width=16cm]{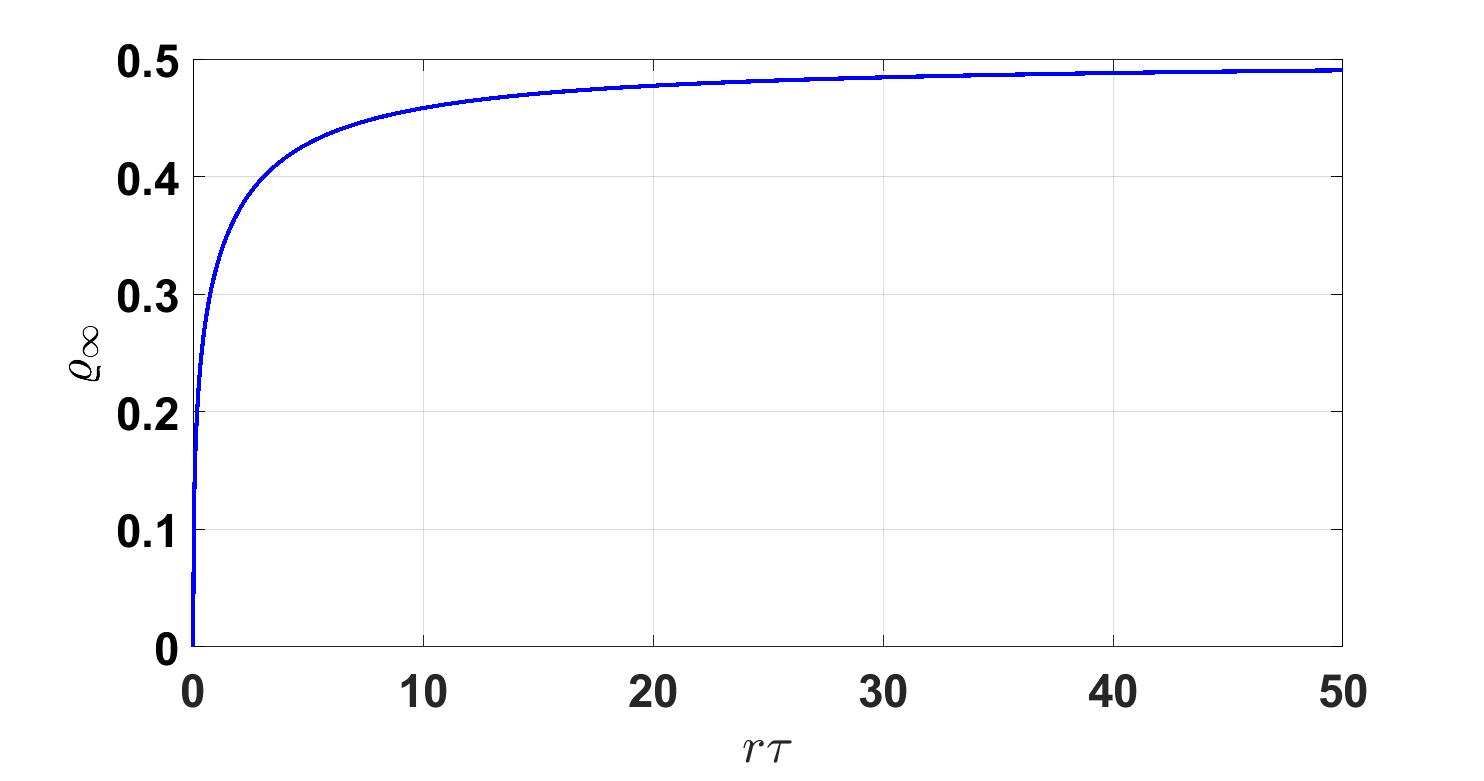} 
\caption{{\bf{The steady-state density of domain walls in the one-dimensional model as a function of the quantity $r\tau$.}} The initial condition consists of independent  undecided voters.}
 \label{figDomainWalls}
\end{center}
\end{figure}

\begin{figure}[H]
\begin{center}
\includegraphics[width=16cm]{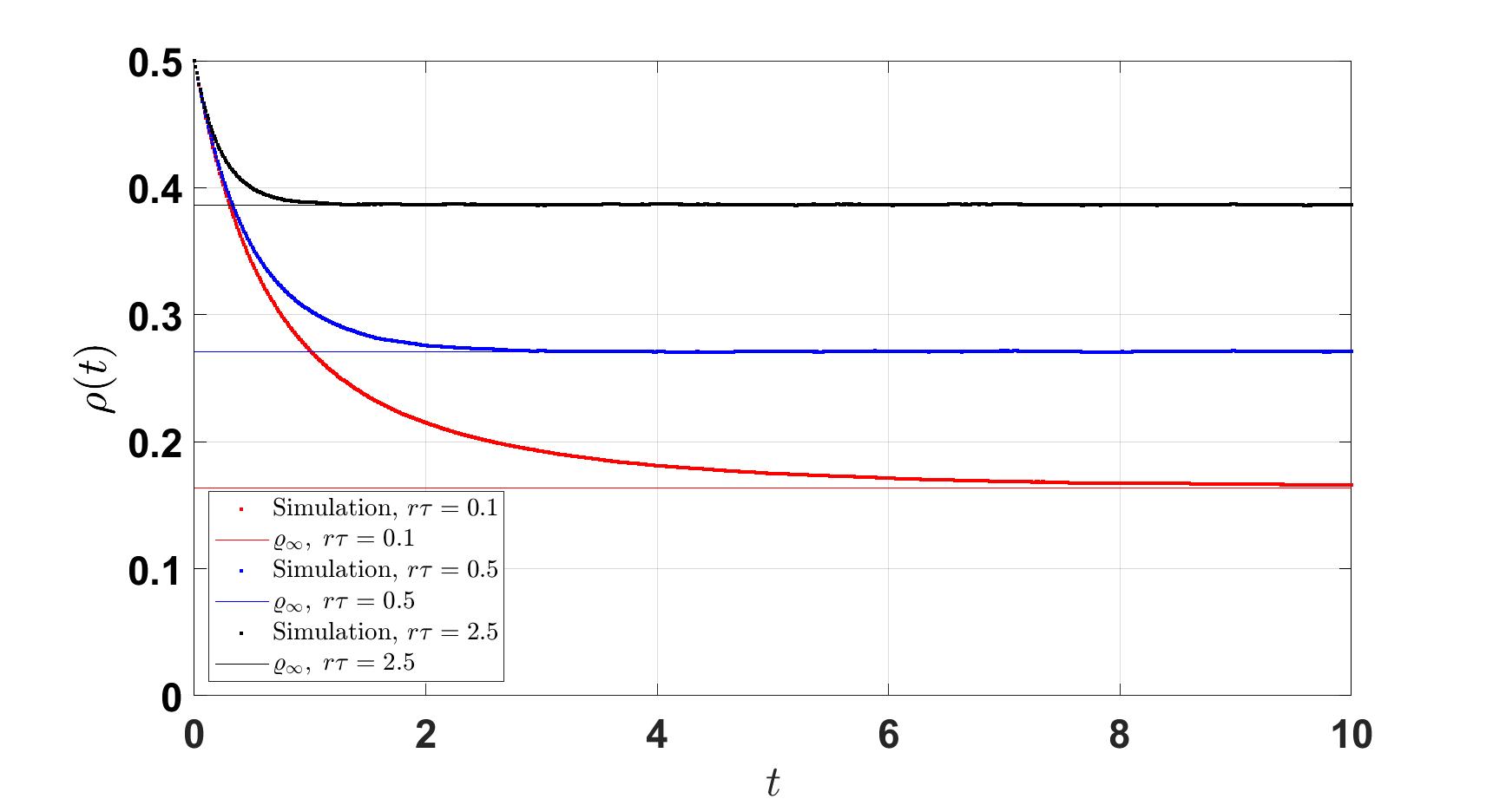} 
\caption{{\bf{Results of numerical simulation of the average density of domain walls.}}  The initial condition consists of independent  undecided voters. The parameter $\tau$ was set to $1$.}
 \label{figDomainWallsSimul}
\end{center}
\end{figure}

 According to 
 Eq. (\ref{rhoDef}) the density of domain walls is related to the two-point function, which yields an expression of  the density of domain walls
 $\varrho$ in terms of the value $\mathcal{G}(a,t)$ :
\begin{equation}\label{varrhoAndGcal}
\begin{split}
 \varrho(t) &= \frac{1}{2}\left(  1 - G( na, (n+1)a, t)  \right) = \frac{1}{2}\left(  1 - \mathcal{G}( a, t)  \right). 
\end{split}
\end{equation}
 From the expression of the two-point function in Eq. (\ref{solGcalReal}) in the particular case $n=1$ we obtain
\begin{equation}
\begin{split}
 1-2 \varrho(t) =& e^{-2(r+\tauInv ) t}I_1( 2\tauInv t) +2r \frac{a}{2\pi} \intB dk \int_0^t du e^{2\alpha(k)(t-u) -ika} \hat{\mathcal{H}}(k,u) \\
 &+ \int_0^t du  e^{-2(r+\tauInv ) (t-u)}I_1( 2\tauInv (t - u))  J(u). 
\end{split}
\end{equation}
 The Laplace transform $\tilde{\varrho}$ of the density of domain walls is therefore related to the  Laplace transform of the current $J$ by
\begin{equation}\label{spotAbove}
 \frac{1}{s} - 2\tilde{\varrho}( s ) =  \tilde{\varphi}_1(s) +2r \Psi_1(s) + \tilde{\varphi}_1(s)  \tilde{J}(s).
\end{equation}
Combining with the value of $\tilde{J}(s)$ obtained in Eq. (\ref{exprJs}) yields
\begin{equation}\label{spotvarrho}
\begin{split}
\tilde{\varrho}( s ) =\frac{1}{2}\left(  \frac{1}{s} - 2r \Psi_1(s) + 2r \frac{\tilde{\varphi}_1(s)}{\tilde{\varphi}_0(s)}\Psi_0(s) - \frac{1}{s} \frac{\tilde{\varphi}_1(s)}{\tilde{\varphi}_0(s)} \right).
\end{split} 
\end{equation}
 Applying the final-value theorem,
 we obtain  the steady-state value $\varrho_\infty$ of the density  of domain walls for an initial state with undecided voters:
 \begin{equation}\label{spot}
\begin{split}
\varrho_\infty &=  \underset{s\to 0}{\lim}\left( s \tilde{\varrho}( s ) \right)\\
&= \frac{1}{2}\underset{s\to 0}{\lim}\left(  1 - 2r s\Psi_1(s) + 2rs  \frac{\tilde{\varphi}_1(s)}{\tilde{\varphi}_0(s)}\Psi_0(s) - \frac{\tilde{\varphi}_1(s)}{\tilde{\varphi}_0(s)} \right),
\end{split} 
\end{equation}
 where we  used again the notations introduced in Eq. (\ref{varPhiDef}). 
 The quantities $\Psi_0(s)$ and $\Psi_1(s)$ and the needed Laplace transforms are worked out in Appendix \ref{AppLap1d} (Eqs (\ref{limphi10},\ref{limPsin})). Substitution yields:
\begin{equation}\label{rhoExpr}
\begin{split}
  \varrho_\infty = 
\frac{1}{2}&\left[   1 -  \frac{(r\tau)^2}{2\pi} \int_{-\pi}^\pi  dx \frac{\cos( x )}{(- \cos( x ) + r\tau +1)^2}\right.\\
& \left. - \frac{1}{ r\tau+1 +\sqrt{(r\tau+1)^2 - 1 }}\left( 1-  \frac{(r\tau)^2}{2\pi} \int_{-\pi}^\pi  dx \frac{1}{(- \cos( x ) + r\tau +1)^2}\right) \right].
\end{split}
\end{equation}
 In dimension one, the above result is the solution
  of Problem (II$_{(r)}$) defined in the introduction. 
 The density $\varrho_\infty$ is plotted  as a function of $r\tau$ on Fig. \ref{figDomainWalls}. Consensus is not achieved in 
  the voter model under resetting in one dimension. However, the steady-state density 
   goes to zero  when the  resetting rate goes to zero. This limit corresponds to the ordinary voter model model, which  is known to lead to a consensus in  dimension one \cite{frachebourg1996exact}.  The vertical tangent at the origin corresponds to the equivalent
\begin{equation}\label{equivRho}
\begin{split}
  \varrho_\infty \underset{r\tau \ll 1}{\sim}  \sqrt{\frac{r\tau}{2}},
\end{split}
\end{equation}
 which is worked out in Appendix \ref{completeness}. 
On the other hand, in the limit of large resetting rate, the density $\varrho_\infty$ goes to $1/2$, which is 
 the average density of domain walls in a system of undecided voters. Direct
 numerical simulations of the evolution of the density of domain walls are shown together with the predicted steady-state value on Fig. \ref{figDomainWallsSimul}. \\

\section{Generalization to higher dimension}\label{higherDim}

 Consider the voter model on a hypercubic $d$-dimensional lattice, with lattice spacing $a$.  Let us pick an orthonormal basis $(\mathbf{e_1},\dots,\mathbf{e_d})$ of 
 ${\mathbb{R}}^d$. There is a voter at every vertex in the lattice, carrying a binary opinion.
 At time $t$, the voter at the site $\mathbf{x}=a \mathbf{n}=  \sum_i a n_i \mathbf{e_i} $ (where $\mathbf{n}$ is an element of $\mathbb{Z}^d$)  is in the opinion  state  $s(\mathbf{x},t)$ in $\{ -1, +1\}$.\\

The flipping rule is adapted from the one-dimensional case (Eq. (\ref{wDef})) as follows. The transition rate at site $\mathbf{x}$ consists of two contributions. The first one is the flipping rate $W$ of the ordinary voter model in dimension $d$, which is the fraction of disagreeing neighbors, with a  factor of $\tau^{-1}$:
\begin{equation}
   W(\mathbf{x},t) :=\frac{\tau^{-1}}{2}\left[1 - \frac{1}{2d}s(\mathbf{x},t)
\sumd [  s(\mathbf{x}+\mathbf{e_i},t) + s(\mathbf{x}-\mathbf{e_i},t) ] \right],\;\;\;\;\;\;\;\;\;\hspace{1cm}(\mathbf{x}\in a\mathbb{Z}^d).
\end{equation}
  The only difference with Eq. (\ref{WDef}), apart from the vector-valued spatial variables, is the factor of $d^{-1}$ in front of the sum over nearest neighbors, which ensures that the maximum flipping rate is $\tau^{-1}$ (achieved in configurations in which all of the $2d$ neighbors of the site $\mathbf{x}$ are occupied by voters with opinion state $-s(\mathbf{x},t)$). This can immediately be checked by using the constraint $s( \mathbf{x},t)^2=1$. 
 The second contribution corresponds to resetting. It is  directly  obtained from Eq. (\ref{RDef}) by substituting vector-valued space variables to the positions of the sites:
\begin{equation}
   R(\mathbf{x},t) := \frac{r}{2}\left(1 - s(\mathbf{x},0)s(\mathbf{x},t)\right),\;\;\;\;\;\;\;\;\;\hspace{1cm}(\mathbf{x}\in a\mathbb{Z}^d).
\end{equation}
The resetting times are Poisson-distributed with intensity $r$, at each resetting time the opinion state of the voter at site $\mathbf{x}$  reverts to its initial value $s(\mathbf{x},0)$ if $s(\mathbf{x},t)\neq s(\mathbf{x},0)$.\\

The transition rate $w$ of the voter model under stochastic resetting in dimension $d$ is therefore expressed as
\begin{equation}\label{wDefd}
 \begin{split}
   w(\mathbf{x},t) :=& W( \mathbf{x},t) + R( \mathbf{x},t)\\  
=& \frac{1}{2}\left[\tau^{-1}\left( 1- \frac{1}{2d}s(\mathbf{x},t)
\sumd [  s(\mathbf{x}+\mathbf{e_i},t) + s(\mathbf{x}-\mathbf{e_i},t) ]\right)
 + r(1 - s(\mathbf{x},0)s(\mathbf{x},t))\right].
\end{split}
\end{equation}
  To study the evolution of the one- and two-point functions of the voter model in dimension $d$, let us adapt the reasoning of the previous section.\\

\subsection{Average opinion state}
 The average opinion state at site  $\mathbf{x}$ and time $t$ is again  denoted by $S$, and defined as the ensemble average 
\begin{equation}
 S( \mathbf{x},t) := \langle s(\mathbf{x},t) \rangle.
\end{equation}
The opinion state at site $\mathbf{x}$ changes by $-2s(\mathbf{x},t)$ when the voter at site $\mathbf{x}$ flips its opinion state.
 This induces the following evolution equation for the average opinion state (generalizing Eq. (\ref{evolMag})):
\begin{equation}\label{evolMagd}
\begin{split}
\frac{\partial S(\mathbf{x},t)}{\partial t} =& - 2 \langle s(\mathbf{x},t) w(\mathbf{x},t) \rangle\\
=& \tau^{-1}\left(-  S(\mathbf{x},t) + 
 \frac{1}{2d}\sum_{i=1}^d[S(\mathbf{x}-a\mathbf{e_i}, t) + S(\mathbf{x}+a\mathbf{e_i},t)] \right)\\
  &+r\left( S(\mathbf{x},0)- S(\mathbf{x},t) \right),\;\;\;\;\;\;\;\;\;\;\shoveright{ (\mathbf{x}\in a{\mathbb{Z}}^d)}.
\end{split}
\end{equation}

 We are instructed to solve this equation with an initial condition consisting of a given 
 function $S_0$ defined on the hypercubic lattice:
\begin{equation}\label{initSd}
 S(\mathbf{x},0) = S_0( \mathbf{x}),\;\;\;\;\;\;\;\;\;\;\;\shoveright{ (\mathbf{x}\in a{\mathbb{Z}}^d).}
\end{equation}

In dimension $d$ the Fourier transform of a function $f$ of the position in the hypercubic lattice 
 (and possibly other variables) is defined as
\begin{equation}
 \hat{f}(\mathbf{k}):= \sum_{\mathbf{n}\in \mathbb{Z}^d} f( \mathbf{n}a) e^{i\mathbf{k}\cdot \mathbf{n}a},\;\;\;\;\;\;\;\;\;\;\;\;\shoveright{\mathbf{k}\in \mathbb{R}^d.}
\end{equation}
The Fourier transform is  inverted by integrating over the first Brillouin zone $[-\pi/a, \pi/a]^d$:
\begin{equation}
f( \mathbf{n}a) = \left(\prod_{i=1}^d  \int_{-\frac{\pi}{a}}^{\frac{\pi}{a}} \frac{a}{2\pi} dk_i\right)\hat{f}(\mathbf{k}) e^{-i \mathbf{k}\cdot \mathbf{n} a} ,\;\;\;\;\;\;\;\;\;\;\;\;\;\;\;\;\shoveright{(\mathbf{n}\in \mathbb{Z}^d).}
\end{equation}
The expressions given in Eqs (\ref{FourierDef},\ref{FourierInv}) are recovered if $d=1$.\\

The Fourier transform of Eqs (\ref{evolMagd},\ref{initSd}) reads
\begin{equation}\label{evolMagFourierd}
\begin{split}
\frac{\partial \hat{S}(\mathbf{k},t)}{\partial t} &= \left[ - (r + \tau^{-1})+\frac{1}{\tau d}\sum_{i=1}^d\cos(k_ia)  \right]  \hat{S}(\mathbf{k},t)+r \hat{S_0}(\mathbf{k}),\\
\hat{S}(\mathbf{k},0 ) &= \hat{S_0}(\mathbf{k}).
\end{split}
\end{equation}
This initial-value problem is readily solved for any vector $\mathbf{k}$ in $\mathbb{R}^d$:
\begin{equation}
\begin{split}
 \hat{S}(\mathbf{k},t) = \exp &\left( [- (r+\tau^{-1}) + \frac{1}{\tau d}\sum_{i=1}^d\cos( k_i a)]t \right)  \hat{S_0}( \mathbf{k}) \\
&+r \int_{0}^t du\,  \hat{S_0}( \mathbf{k}) \exp\left( [- (r+\tau^{-1}) + \frac{1}{\tau d}\sum_{i=1}^d\cos( k_i a)] u\right).
\end{split}
\end{equation}
Fourier inversion yields the opinion state profile at position $\mathbf{n}a$ and time $t$ as
\begin{equation}\label{explicitSd}
\begin{split}
 S(\mathbf{n}a,t)   =&\left( \prod_{j=1}^d    \int_{-\frac{\pi}{a}}^{\frac{\pi}{a}}  \frac{a}{2\pi}  dk_j\right) \exp\left( \left[- (r+\tau^{-1}) + \frac{1}{\tau d}\sum_{j=1}^d\cos( k_j a)\right]t -i \mathbf{k}
\cdot \mathbf{n} a \right)  \hat{S_0}(\mathbf{k}) \\
 &+ r \int_{0}^t du   \left( \prod_{j=1}^d     \int_{-\frac{\pi}{a}}^{\frac{\pi}{a}} \frac{a}{2\pi}  dk_j\right)   \hat{S_0}(\mathbf{k}) 
                      \exp\left(  \left[- (r+\tau^{-1}) + \frac{1}{\tau d}\sum_{j=1}^d\cos( k_j a)\right] u -i \mathbf{k}\cdot \mathbf{n} a \right).
\end{split}
\end{equation}
Let us use the identity  
 $e^{(\tau d)^{-1} t\cos( k_ja)} = \sum_{ m \in \mathbb{Z}} I_m((\tau d)^{-1} t) e^{imk_ja}$
 every  $j$ in $[1..d]$ (which is the generating function of the modified Bessel functions of the first kind, given in Eq. (\ref{BesselGener})). Taking the same steps as in Eq. (\ref{convolExact}), we obtain 
\begin{equation}\label{explicitSdNext}
\begin{split}
 S(\mathbf{n}a,t)   =& \exp\left( - (r+\tau^{-1})t \right) \left( \prod_{j=1}^d    \int_{-\frac{\pi}{a}}^{\frac{\pi}{a}}\frac{a}{2\pi}  dk_j\right) \hat{S_0}(\mathbf{k}) \prod_{l=1}^d \exp\left(   \frac{t}{\tau d} \cos(k_l a ) -ia
 k_l n_l \right) \\
 &+ r \int_{0}^t du  \exp\left( - (r+\tau^{-1})u\right) \left( \prod_{j=1}^d     \int_{-\frac{\pi}{a}}^{\frac{\pi}{a}}\frac{a}{2\pi}  dk_j\right)   \hat{S_0}(\mathbf{k}) \prod_{l=1}^d \exp\left(   \frac{u}{\tau d} \cos(k_l a ) -ia k_l n_l \right)  \\
=& \exp\left( - (r+\tau^{-1})t \right) \left( \prod_{j=1}^d    \int_{-\frac{\pi}{a}}^{\frac{\pi}{a}}\frac{a}{2\pi}  dk_j\right) \hat{S_0}(\mathbf{k})
 \prod_{l=1}^d \sum_{m_l \in \mathbb{Z}} I_{m_l}\left((\tau d)^{-1} t \right) e^{im_l k_la} e^{-ia
 k_l n_l}\\
 &+ r \int_{0}^t du  \exp\left( - (r+\tau^{-1})u\right) \left( \prod_{j=1}^d \int_{-\frac{\pi}{a}}^{\frac{\pi}{a}} \frac{a}{2\pi} dk_j\right)   \hat{S_0}(\mathbf{k}) \prod_{l=1}^d  
  \sum_{m_l \in \mathbb{Z}} I_{m_l}\left((\tau d)^{-1} u \right) e^{im_l k_la} e^{-ia
 k_l n_l}\\
=& \exp\left( - (r+\tau^{-1})t \right) 
 \sum_{\mathbf{m} \in \mathbb{Z}^d} I_{\mathbf{m}}\left((\tau d)^{-1} t \right) S_0( \mathbf{n}a
 -\mathbf{m}a)\\
 &+ r \int_{0}^t du  \exp\left( - (r+\tau^{-1})u\right) \sum_{\mathbf{m} \in \mathbb{Z}^d} I_{\mathbf{m}}\left((\tau d)^{-1} t \right) S_0( \mathbf{n}a-\mathbf{m}a),
\;\;\;\;\;\;\;\;\;\;\;    (\mathbf{n} \in {\mathbb{Z}}^d),
\end{split}
\end{equation}
where the multi-index modified Bessel function $I_{\mathbf{m}}$  is defined
 as:
\begin{equation}
I_{\mathbf{m}}( t ) = \prod_{j=1}^d I_{m_j}(t),
\;\;\;\;\;\;\;\;\;\;\;    (\mathbf{m} \in {\mathbb{Z}}^d).
\end{equation}

 The average opinion state at  time $t$ is therefore the convolution product of the initial average opinion state and a time-dependent kernel denoted again by $\mathcal{K}_t$ (which generalizes  Eq. (\ref{KDef}) to dimension $d$):
 \begin{equation}\label{KDefd}
\begin{split}
   S( \mathbf{n} a,t) =&\sum_{\mathbf{m}\in \mathbb{Z}^d} {\mathcal{K}}_t(\mathbf{m}a) S_0( \mathbf{n}a-\mathbf{m}a),\\
{\mathrm{with}}\;\;\;\;\;{\mathcal{K}}_t(\mathbf{m}a) :=&
\exp\left( - (r+\tau^{-1})t \right)  I_{\mathbf{m}}\left((\tau d)^{-1} t \right)\\
 &+ r \int_{0}^t du  \exp\left( - (r+\tau^{-1})u\right)  I_{\mathbf{m}}\left((\tau d)^{-1} t \right),
\;\;\;\;\;\;\;\;\;\;\;  (\mathbf{m} \in {\mathbb{Z}}^d).
 \end{split}
\end{equation}
 Taking the Fourier transform of the discrete convolution product yields the product 
\begin{equation}
\begin{split}
\hat{S}(\mathbf{k},t) =& \hat{\mathcal{K}}_t(\mathbf{k}) \hat{S_0}(\mathbf{k}).\\
\end{split}
\end{equation}
 Moreover, we can read off the Fourier transforms $\hat{S}(\mathbf{k},t)$ 
 from the integrand on the r.h.s. of Eq. (\ref{explicitSd}). We deduce the Fourier transform of the kernel and its large-time limit:
\begin{equation}\label{KInfComp}
\begin{split}
\mathcal{K}_t( \mathbf{k}) =& \exp\left( \left[- (r+\tau^{-1}) + \frac{1}{\tau d}\sum_{j=1}^d\cos( k_j a)\right]t  \right)   \\
 &+ r \int_{0}^t du 
                      \exp\left(  \left[- (r+\tau^{-1}) + \frac{1}{\tau d}\sum_{j=1}^d\cos( k_j a)\right] u\right)\\
 =& \frac{r+\tau^{-1}( 1- d^{-1}\sum_{j=1}^d\cos(k_ja)) e^{-[ r +\tau^{-1}(1- d^{-1}\sum_{j=1}^d\cos(k_ja) )] t}}{r +\tau^{-1}(1-  d^{-1}\sum_{j=1}^d\cos(k_ja))},\\
\mathcal{K}_\infty( \mathbf{k}) = & \frac{r}{r +\tau^{-1}(1-  d^{-1}\sum_{j=1}^d\cos(k_ja))},
                    \;\;\;\;\;\;\;\;\;\;\;\;(\mathbf{k} \in {\mathbb{R}}^d).
\end{split}
\end{equation}
 We have thus generalized Eqs (\ref{KFour},\ref{KFourInf}).
 Let us derive an upper bound on $|S(\mathbf{n}a,t)-S(\mathbf{n}a,\infty)|$ by generalizing Eq. (\ref{upperBound1d}):
\begin{equation}\label{upperBound}
\begin{split}
&\left|\hat{S}(\mathbf{n}a,t) - \hat{S}(\mathbf{n}a,\infty) \right|= \left|  \left( \prod_{j=1}^d    
\int_{-\frac{\pi}{a}}^{\frac{\pi}{a}}\frac{a}{2\pi}  dk_j\right) 
\left[ 
\hat{S}_0(\mathbf{k})
\left( \hat{\mathcal{K}}_t(\mathbf{k})- \hat{\mathcal{K}}_\infty(\mathbf{k})\right)
\right]
e^{-i \mathbf{n} \cdot\mathbf{k}a}\right|\\
=&\left|
\left( \prod_{j=1}^d    \int_{-\frac{\pi}{a}}^{\frac{\pi}{a}}\frac{a}{2\pi}  dk_j
\right) 
\left[ \hat{S}_0(\mathbf{k})
\frac{\tau^{-1}( 1- d^{-1}\sum_{j=1}^d\cos(k_ja)) e^{-[ r +\tau^{-1}(1- d^{-1}\sum_{j=1}^d\cos(k_ja) )] t}}{r +\tau^{-1}(1-  d^{-1}\sum_{j=1}^d\cos(k_ja))}
\right] e^{-i \mathbf{n} \cdot\mathbf{k}a}\right|\\
=& e^{-rt}\left|\left( \prod_{j=1}^d    \int_{-\frac{\pi}{a}}^{\frac{\pi}{a}}\frac{a}{2\pi}  dk_j\right) \hat{S}_0(\mathbf{k})
\frac{\tau^{-1}( 1- d^{-1}\sum_{j=1}^d\cos(k_ja)) e^{-\tau^{-1}(1-d^{-1}\sum_{j=1}^d\cos(k_ja) ) t}}{r +\tau^{-1}(1-d^{-1}\sum_{j=1}^d\cos(k_ja) )}
 e^{-i \mathbf{n} \cdot\mathbf{k}a}\right|\\
\leq &e^{-rt}\left( \prod_{j=1}^d    \int_{-\frac{\pi}{a}}^{\frac{\pi}{a}}\frac{a}{2\pi}  dk_j\right) \left|\hat{S}_0(\mathbf{k})e^{-\tau^{-1}(1-d^{-1}\sum_{j=1}^d\cos(k_ja) ) t}
\frac{\tau^{-1}( 1- d^{-1}\sum_{j=1}^d\cos(k_ja))}{r +\tau^{-1}(1-d^{-1}\sum_{j=1}^d\cos(k_ja) )}\right|\\
\leq &e^{-rt}\left( \prod_{j=1}^d    \int_{-\frac{\pi}{a}}^{\frac{\pi}{a}}\frac{a}{2\pi}  dk_j\right) \left|\hat{S}_0(\mathbf{k})\right| e^{-\tau^{-1}(1-d^{-1}\sum_{j=1}^d\cos(k_ja) ) t},\;\;\;\;\;\;\;\;\;\;\; (\mathbf{n} \in {\mathbb{Z}}^d).
\end{split}
\end{equation}
 Again the upper bound on the r.h.s. is independent of the position  $\mathbf{n}a$, and goes to zero if $\hat{S}_0$ is integrable on the first Brillouin zone.\\


As an example, let us work out the solution  for an initial condition corresponding to undecided voters, except for a single decided voter at the origin with positive opinion. At every site in the lattice it coincides with the time-dependent kernel $\mathcal{K}_t$. Indeed, working out the discrete convolution in Eq. (\ref{KDefd}) yields 
\begin{equation}
\begin{split}
 S_0( \mathbf{n} a ) =& \delta_{\mathbf{n},\mathbf{0}},\\
 S( \mathbf{n}a, t) =& {\mathcal{K}}_t(\mathbf{n}a )\\
                            =& \exp\left( - (r+\tau^{-1})t \right) 
   I_{\mathbf{n}}\left((\tau d)^{-1} t \right) \\
 &+ r \int_{0}^t du  \exp\left( - (r+\tau^{-1})u\right)  I_{\mathbf{n}}\left((\tau d)^{-1} u \right),\;\;\;\;\;\;    \mathbf{n} \in {\mathbb{Z}}^d,
\end{split}
\end{equation}
which generalizes Eq. (\ref{singleDecided}). 
 The average opinion state is equal to the probability of presence  of a diffusive random walker  on a $d$-dimensional hypercubic lattice starting from the origin, and reset to its initial position at  times generated by a Poisson process of intensity $r$.\\


 The average opinion state in the steady state of the system is the large-time limit of the above equation. It is given in terms of the Laplace transform\footnote{The Laplace transform of functions of time and other variables has been defined in Eq. (\ref{LaplaceTransformDef}).} of the multi-index modified Bessel functions:
\begin{equation}\label{SnaInfd}
\begin{split}
 S( \mathbf{n}a, \infty) =&  r \int_{0}^\infty du  \exp\left( - (r+\tau^{-1})u\right)  I_{\mathbf{n}}\left((\tau d)^{-1} u \right)\\
=& r \tau \int_{0}^\infty dv \exp\left( - (r\tau+1) v\right)  I_{\mathbf{n}}\left( d^{-1}v\right)\\
=& r \tau \int_0^\infty dv \exp\left( - (r\tau+1) v\right)  \left( \prod_{j=1}^d   \intP \frac{dq_j}{2\pi}\right) \exp\left(  \frac{1}{d}\sum_{j=1}^d v\cos( q_j ) \right) e^{-i\sum_{j=1}^d n_j q_j}\\
=& r \tau  \left( \prod_{j=1}^d  \intP \frac{dq_j}{2\pi}  \right)  \int_0^\infty dv  \exp\left( \left[ -( r\tau + 1 ) + \frac{1}{d}\sum_{j=1}^d \cos( q_j )  \right] v\right) e^{-i\sum_{j=1}^d n_j q_j}\\
=& r\tau  \left( \prod_{j=1}^d  \intP \frac{dq_j}{2\pi}  \right) \frac{ \cos(n_j q_j)}{r\tau + 1 -\frac{1}{d}\sum_{i=1}^d \cos(q_j )},
\end{split}
\end{equation}
  where in the third step we inserted the generating function of modified Bessel functions defined in Eq. (\ref{BesselGener}). Moreover, we used parity to replace $e^{in_j q_j}$ with $\cos( n_j q_j)$ in the integrand. We may recover this expression by using  the Fourier representation of $\mathcal{K}_t(\mathbf{n}a)$ and using Eq. (\ref{KInfComp}). Indeed,
  \begin{equation}
  \begin{split}
    S(\mathbf{n}a,\infty) =& \mathcal{K}_\infty(\mathbf{n}a) = \left( \prod_{j=1}^d  \intP \frac{dk_j}{2\pi}  \right) \mathcal{K}_\infty(\mathbf{n}a) e^{-i\sum_{j=1}^d k_j n_j}\\
    =& \left( \prod_{j=1}^d  \intP \frac{a}{2\pi} dk_j \right)
     \frac{r}{r +\tau^{-1}(1-  d^{-1}\sum_{j=1}^d\cos(k_ja))} e^{-i\sum_{j=1}^d k_j n_j}\\
    =& r\tau   \left( \prod_{j=1}^d  \intP \frac{dq_j}{2\pi}  \right)
     \frac{cos(n q_j)}{ r\tau + 1 - \frac{1}{d}\sum_{j=1}^d\cos(q_j)}.
  \end{split}
  \end{equation}

  The stationary state of the average opinion state is therefore (up to a factor or $r\tau$), the Laplace transform at $r\tau + 1$ of the probability of presence of a random diffusive walker starting from the origin on a hypercubic lattice.\\

 Consider the average opinion state at the origin in the steady state.  Let us can rewrite its expression obtained in Eq. (\ref{SnaInfd}) to obtain the solution of Problem (I$_{(r)}$) defined in the Introduction:
\begin{equation}
\begin{split}
 S( \mathbf{0}, \infty) =& 2r\tau  \mathcal{I}_d( 2r\tau),
\end{split}
\end{equation}
 with the notation (borrowed from Chapter 8 of \cite{kineticView}):
\begin{equation}\label{Id}
{\mathcal{I}}_d(\epsilon) :=  \left(\prod_{i=1}^d  \int_{-\pi}^\pi \frac{dq_i}{2\pi} \right) \frac{1}{ \epsilon + \frac{2}{ d}  \sum_{i=1}^d\left( 1 - \cos(q_i ) \right) }.
\end{equation}

 Let us check consistency with the results reported in Section \ref{QOI}. In dimension one,   Eq. (\ref{lambdaDef}) yield $S(0,\infty) = r\tau \tilde{I}_0( r\tau + 1)$. Using the Fourier representation of the Bessel function $I_0$, we calculate:
\begin{equation}
\begin{split}
 r\tau \tilde{I}_0( r\tau + 1) &= 
r\tau \int_0^{\infty} du e^{-( r\tau +1) u} \int_{-\pi}^\pi\frac{dq}{2\pi} e^{u \cos(q)}\\
&= r\tau \int_{-\pi}^\pi\frac{dq}{2\pi} \frac{1}{ r\tau + 1 - \cos(q)} = 2r\tau \int_{-\pi}^\pi\frac{dq}{2\pi} \frac{1}{ 2r\tau +  2( 1 - \cos(q) )}\\
 &= 2r\tau  \mathcal{I}_1( 2r\tau).
\end{split}
\end{equation}
  The general result obtained in Eq. (\ref{SnaInfd}) is therefore consistent with the solution of the model in dimension one. In the limit of zero resetting rate, the voter model under stochastic resetting reduces to the ordinary voter model and the opinion of the initially-decided voter should relax to the undecided state of the rest of the population. On the other hand, the behavior of the average opinion state in the steady state  at low resetting rate depends on the behavior of  $\mathcal{I}_d(2r\tau)$ when $r\ll \tau^{-1}$. 
  This behavior depends on the dimension and is worked out in the appendices: $\mathcal{I}_d(0)$ is finite if $d\geq 3$. In particular, the quantity $\mathcal{I}_3(0)$ is known exactly, as one of the  Watson integrals (see Eq. (\ref{Watson3})). 
  We obtain the equivalent $\mathcal{I}_2(2r\tau) \underset{r\tau\ll 1}{\sim}  -(2\pi)^{-1}\ln( r\tau )$ (see Eq. (\ref{I2Dev})). Moreover, the equivalent $  \mathcal{I}_{1}(2r\tau) \underset{r\tau\ll 1}{\sim} (2\sqrt{2r\tau})^{-1}$ (see Eq. (\ref{I1Dev})) allows to recover the result reported in Eq. (\ref{equiv1dS}). Substituting into Eq. (\ref{SnaInfd}) yields the equivalents
 \begin{equation}
 S( \mathbf{0}, \infty) \underset{r\ll \tau^{-1}}{\sim}
\begin{cases}
  2r\tau \mathcal{I}_d(0) ,\;\;\;\;\;\;\;\;\;&(d\geq 3),\\
 -\frac{r\tau}{\pi}\ln( r\tau ),\;\;\;\;\;\;\;\;\;&(d=2),\\
 \sqrt{\frac{r\tau}{2}},   \;\;\;\;\;\;\;\;\;\;&(d= 1).\\
\end{cases}
\end{equation}  
In particular, the average opinion state at the origin has a finite susceptibility $\frac{\delta S(\mathbf{0},\infty)}{\delta r}|_{r=0}$ if and only if $d\geq 3$.\\ 


{\bf{Continuum limit.}} Let us take again the continuum limit by sending the lattice spacing $a$ and the time $\tau$ to zero, with a fixed diffusion coefficient $D = a^2/(2\tau)$. Consider a position $\mathbf{x}$ in the lattice, with a large norm on the scale of the lattice spacing:
\begin{equation}
  \mathbf{x} = {\mathbf{n}}a,\;\;\;\;\;\;\;\;{\mathrm{with}}\;\;\;\;\;\;\;|| {\mathbf{n}}||  \gg 1.
\end{equation}
 Changing integration variables to $t:=\tau v$ and $k_j:=a^{-1} q_j,\;(1\leq j\leq d)$ in the third row of Eq.  (\ref{SnaInfd}), we obtain the density of opinion state in the diffusive limit as
\begin{equation}
\begin{split}
a^{-d}S( a\mathbf{n},\infty ) =& r\int_0^\infty dw \left( \prod_{j=1}^d  \intB \frac{dk_j}{2\pi}  \right)\exp\left( -rw + \frac{w}{\tau d}\sum_{j=1}^d( -1 + \cos( a k_j)  )- i \mathbf{k}\cdot\mathbf{x}    \right)\\
\underset{\substack{a\to 0, \;||\mathbf{n}||\to \infty,\\
   \mathbf{x}=\mathbf{n}a, \;D = \frac{a^2}{2\tau}}}\sim &  r  \int_0^\infty dw e^{-rw}
\left( \prod_{j=1}^d  \int_{-\infty}^\infty \frac{dk_j}{2\pi}  \right) \exp\left( - \frac{wD}{d} \mathbf{k}\cdot\mathbf{k}  - i \mathbf{k}\cdot\mathbf{x}    \right)\\
 =& r \int_0^\infty dw e^{-rw}  \frac{1}{(4\pi D w)^{\frac{d}{2}}}\exp\left( -\frac{|| \mathbf{x}||^2}{4Dw} \right),
\end{split}
\end{equation}
 where in the last step we have worked out a Gaussian integral. We recognize the steady-state probability density $p^\ast( \mathbf{x})$ of a diffusive random walker under stochastic resetting to the origin in ${\mathbb{R}}^d$ \cite{evans2014diffusion}.
 The relevant Laplace transform can be expressed in terms of the modified Bessel function of the second kind $K_{1-d/2}$ (which appears because of the identity $\int_0^\infty dt\, t^{\nu -1} e^{-\frac{\beta}{t}-\gamma t} = 2
 (\gamma^{-1}\beta)^{\frac{\nu}{2}} K_\nu( 2\sqrt{\beta\gamma})$). We just quote the result from  \cite{evans2014diffusion}:
\begin{equation}
p^\ast(\mathbf{x}) = \left( \frac{r}{2\pi D} \right)\left(  \sqrt\frac{r}{D}|| \mathbf{x}||\right)^{1-\frac{d}{2}} K_{1-\frac{d}{2}}( \sqrt{D^{-1}r}|| \mathbf{x}||).
\end{equation}
  The identity $K_{1/2}(y) = \sqrt{(2y)^{-1}\pi}e^{-y}$ allows to recover Eq. (\ref{continuum1d}) in dimension $d=1$. The  density  $p^\ast(\mathbf{x})$ diverges at the origin in dimension $d\geq 2$, even though it  integrates to $1$ over $\mathbb{R}^d$ (see Section 2.6 and Fig. 3 in \cite{topical} for  a review and a plot).\\

\subsection{Two-point function and density of domain walls}
 To calculate density of domain walls, let us consider the two-point function of the model in dimension $d$:
\begin{equation}
G( \mathbf{n}a, \mathbf{m}a,t):= \langle s( \mathbf{n}a,t) s( \mathbf{m}a,t)\rangle,\;\;\;\;\;\;\;\;\;\;\;\;\;(\mathbf{m},\mathbf{n}\in {\mathbb{Z}}^d).
\end{equation}

Let us derive its evolution equation. When the opinion state $s(\mathbf{x},t)$ is flipped,
 its value is shifted by $-2s(\mathbf{x},t)$. 
 If $\mathbf{n}\neq \mathbf{m}$, the transition rates at sites $\mathbf{n}a$ and $\mathbf{m}a$ induce
\begin{equation}\label{aboveG}
\frac{\partial}{\partial t}G( \mathbf{n}a, \mathbf{m}a,t) = -2 \langle s(\mathbf{n}a,t) s(\mathbf{n}a,t)[ w(\mathbf{n}a,t)  + w(\mathbf{m}a,t)]   \rangle,\;\;\;\;\;\;\;\;\;\;(\mathbf{m},\mathbf{n}\in \mathbb{Z}^d, \mathbf{m}\neq\mathbf{n}).\\
\end{equation}
 On the other hand, the  identity $s(\mathbf{x},t)^2=1$ implies that the two-point function at coinciding points is constant:
\begin{equation}
 G( \mathbf{n}a, \mathbf{n}a,t) = 1,\;\;\;\;\;\;\;\;\;\;\;\;(\mathbf{n}\in \mathbb{Z}, t\geq 0).
\end{equation}

 Substituting into Eq. (\ref{aboveG}) the transition rates defined in Eq. (\ref{wDefd}) and using the constraint $s(\mathbf{n}a,t )^2 =1$ yields
\begin{equation}\label{eqGd}
\begin{split}
\frac{\partial}{\partial t}G( \mathbf{n}a, \mathbf{m}a,t) =& -2( r +\tauInv) G( \mathbf{n}a, \mathbf{m}a, t)\\
 + \frac{\tauInv}{d}&\sum_{i=1}^d
\left[ G( \mathbf{n}a - a\mathbf{e_i}, \mathbf{m}a,t) +  G( \mathbf{n}a + a\mathbf{e_i}, \mathbf{m}a,t) \right.\\
&\left.+  
  G( \mathbf{n}a, \mathbf{m}a - a\mathbf{e_i},t) +  G( \mathbf{n}a, \mathbf{m}a - a\mathbf{e_i},t)  \right]\\
&+r [H( \mathbf{n}a, \mathbf{m}a, t )+ H( \mathbf{m}a , \mathbf{n}a, t )],
 \;\;\;\;\;\;(\mathbf{m},\mathbf{n}\in \mathbb{Z}^d, \mathbf{m}\neq\mathbf{n}).
\end{split}
\end{equation}
This equation generalizes Eq. (\ref{evolGH}). We used again the symbol $H$ to denote the two-time two-point function 
\begin{equation}
H( \mathbf{n}a, \mathbf{m}a, t ):= \langle s( \mathbf{n}a, t) s( \mathbf{m}a, 0)\rangle,\;\;\;\;\;\;\;\;\;\;
(\mathbf{m},\mathbf{n}\in \mathbb{Z}^d).
\end{equation}
 The evolution equation of $H$ follows from the flipping rule as
\begin{equation}\label{eqHd}
\begin{split}
 \frac{\partial H}{\partial t}(\mathbf{n}a, \mathbf{m}a,t) =& -2  \langle w(\mathbf{n}a,t) s(\mathbf{n}a,t) s(\mathbf{m}a,0)   \rangle\\
=& - (r+\tauInv) H(\mathbf{n}a, \mathbf{m}a,t) + \frac{\tauInv}{2d} \sum_{i=1 }^d
\left[ H(\mathbf{n}a + \mathbf{e_i}, \mathbf{m}a ,t)  + H(\mathbf{n}a - \mathbf{e_i}, \mathbf{m}a ,t)\right]  \\
&+  r G(\mathbf{n}a, \mathbf{m}a,0),\;\;\;\;\;\;\;\;\;\;(\mathbf{m},\mathbf{n}\in \mathbb{Z}^d),
\end{split}
\end{equation}
which generalizes Eq. (\ref{evolH}). As in dimension one, we notice that the evolution equation of the two-time two-point function (for any fixed value of the second argument), satisfies the same evolution equation 
 as the average opinion state (which in dimension $d$ is Eq. (\ref{evolMagd})).\\

Consider an initial condition with independent undecided voters. 
  The system is again translationally invariant: the values of the functions $G$ and $H$ depend only the separation 
 of the voters (and on time). Hence there exist two functions (denoted again by $\mathcal{G}$ and $\mathcal{H}$) such that
\begin{equation}\label{GHDefd}
\begin{split}
   G( \mathbf{n}a, \mathbf{m}a,t) &=
 \mathcal{G}( (\mathbf{m}-\mathbf{n})a,t ),\\ 
   H(  \mathbf{n}a, \mathbf{m}a,t) &= \mathcal{H}(\mathbf{m}-\mathbf{n})a,t),\;\;\;\;\;\;\;\;(\mathbf{m},\mathbf{n}\in {\mathbb{Z}}^d).
\end{split}
\end{equation}

Let us rewrite Eqs (\ref{eqGd},\ref{eqHd}) in terms of the  functions $\mathcal{G}$ and $\mathcal{H}$:
\begin{equation}\label{myEvolGd}
\begin{split}
 \frac{\partial \mathcal{G}}{\partial t}(\mathbf{n}a,t) =& 
-2( r + \tauInv)\mathcal{G}(\mathbf{n}a,t)\\
 &+\frac{1}{d} \sumd [\mathcal{G}((\mathbf{n}+\mathbf{e_i})a,t)  + \mathcal{G}((\mathbf{n}-\mathbf{e_i})a,t)] + r \mathcal{H}(  \mathbf{n}a,t) +r \mathcal{H}(-  \mathbf{n}a,t),
  \;\;\;\;\;\;\;  ( \mathbf{n} \in \mathbb{Z}^d, \mathbf{n}\neq \mathbf{0}),\\
\end{split}
\end{equation} 
\begin{equation}\label{evolHd}
\begin{split}
 \frac{\partial \mathcal{H}}{\partial t} (\mathbf{n}a,t)=& 
   -( r + \tauInv)\mathcal{H}(\mathbf{n}a,t) \\
&+ \frac{\tauInv}{2d}
 \sum_{i=1}^d  
  [\mathcal{H}((\mathbf{n}+\mathbf{e_i})a,t) +\mathcal{H}((\mathbf{n}-\mathbf{e_i})a,t) ]  + r\mathcal{G}(\mathbf{n}a,0),   \;\;\;\;\;\;\;  (\mathbf{n} \in \mathbb{Z}^d).
\end{split}
\end{equation} 
 The above two equations generalize Eqs (\ref{evolGcal1d},\ref{evolHcal1d}) to dimension $d$. The initial condition  $H(\mathbf{n}a, \mathbf{m}a,0)= \langle s( \mathbf{n}a,0), s( \mathbf{m}a,0) \rangle=
 \delta_{\mathbf{n},\mathbf{m}}$ induces
 $\mathcal{H}(\mathbf{n}a,0) = \delta_{\mathbf{n},\mathbf{0}}$. Hence $\mathcal{H}(\mathbf{n}a,t)$
  is the probability of presence at site $\mathbf{n}a$ and time $t$ of a diffusive random walker on the hypercubic lattice, subjected to  stochastic resetting  to its initial position at the origin. In particular $\mathcal{H}(\mathbf{n}a,t)=\mathcal{H}(-\mathbf{n}a,t)$ for all $\mathbf{n}$ in $\mathbb{Z}^d$, and 
 the last  two terms in Eq. (\ref{myEvolGd}) are equal.\\ 

 Moreover the constraint $s(\mathbf{x},t)^2=1$, which holds for every vertex $\mathbf{x}$ in the lattice,
 induces 
 \begin{equation}\label{constraintGd}
 \mathcal{G}(\mathbf{0},t) =1,\;\;\;\;\;\;\;\;\;(t\geq 0).
\end{equation}

  The initial condition with independent undecided voters reads 
\begin{equation}
\mathcal{G}( \mathbf{n}a, 0 ) = \delta_{\mathbf{n},\mathbf{0}},\;\;\;\;\;\;\;  (\mathbf{n} \in \mathbb{Z}^d).
\end{equation}
  Let us  add a source term $J$ at the origin (an unknown function of time) so that the two-point function 
 $\mathcal{G}$ satisfies the constraint of Eq. (\ref{constraintGd}) at all times:
\begin{equation}\label{evolGSourced}
\begin{split}
\frac{\partial \mathcal{G}}{\partial t}( \mathbf{n}a, t) =& -2( r + \tauInv) \mathcal{G}( \mathbf{n}a, t) 
+ \frac{1}{\tau d} \sum_{i=1}^d [ {\mathcal{G}}( (\mathbf{n}+\mathbf{e_i})a, t) + {\mathcal{G}}( (\mathbf{n}-\mathbf{e_i})a,t) ] \\
 &+ 2r \mathcal{H}( \mathbf{n}a,t) + J(t) \delta_{\mathbf{n},\mathbf{0}}, \;\;\;\;\;\;\;\;\;(\mathbf{n}\in \mathbb{Z}^d),\\
\end{split}
\end{equation} 
 This generalizes  Eq. (\ref{evolGsource}) to dimension $d$.\\

 Taking the Fourier transform yields an ordinary differential equation
\begin{equation}
\frac{\partial \hat{\mathcal{G}}}{\partial t}( \mathbf{k},t) = -2(r +\tauInv) \hat{\mathcal{G}}(\mathbf{k},t) +
 \frac{2}{\tau d} \sum_{j=1}^d\cos( k_j a )  \hat{\mathcal{G}}(\mathbf{k},t) + 2r \hat{\mathcal{H}}(\mathbf{k},t) + J(t),
\end{equation}
together with the initial condition 
$\hat{\mathcal{G}}(\mathbf{k} ,t)=1$.
  This initial-value problem is readily solved with the  as
\begin{equation}\label{soldG}
\begin{split}
\hat{\mathcal{G}}( \mathbf{k},t) &= e^{2 \alpha(\mathbf{k}) t} +2r \int_0^t du e^{2 \alpha(\mathbf{k}) (t-u)} \hat{\mathcal{H}}(k,u)
 + \int_0^t du  e^{2  \alpha(\mathbf{k})(t-u)} J(u),\\
{\mathrm{with}}&\;\;\;\;\;\;\alpha( \mathbf{k} ) :=- (r+\tauInv) + \taudInv\sum_{i=1}^d \cos( k_i a),\;\;\;\;\;\;\;\;\;(\mathbf{k}\in {\mathbb{R}}^d).
\end{split}
\end{equation}
 We need the Fourier transform of the function $\mathcal{H}$. It is obtained
 by taking the Fourier transform of Eq. (\ref{evolHd}) and solving with the initial condition $\hat{\mathcal{H}}(\mathbf{k},0) = 1$:
\begin{equation}\label{soldH}
\begin{split}
\hat{\mathcal{H}}( \mathbf{k}, t ) &=  e^{\alpha(\mathbf{k}) t} + r\int_0^t du e^{  \alpha(\mathbf{k})u}=e^{\alpha(\mathbf{k}) t} +\frac{1}{\alpha(\mathbf{k})}\left( e^{\alpha(\mathbf{k}) t} - 1  \right) ,\;\;\;\;\;\;(\mathbf{k}\in {\mathbb{R}}^d).\\
\end{split}
\end{equation}
Substituting into Eq. (\ref{soldG}) yields
\begin{equation}
\begin{split}
\hat{\mathcal{G}}( \mathbf{k},t) &= e^{2 \alpha(\mathbf{k}) t} +2r \int_0^t du e^{2 \alpha(\mathbf{k}) (t-u)} \hat{\mathcal{H}}(k,u)
 + \int_0^t du  e^{2  \alpha(\mathbf{k})(t-u)} J(u).
\end{split}
\end{equation}
 Inverting the Fourier transform yields the two-point function at time $t$ for initially
 undecided voters in dimension $d$ as
\begin{equation}
\begin{split}
{\mathcal{G}} ( \mathbf{n}a,t) =& 
  \left( \prod_{i=1}^d  \intB  \frac{a}{2\pi} dk_i \right)   e^{-i \mathbf{n} \cdot \mathbf{k}a} \hat{\mathcal{G}}( \mathbf{k},t). \\
\end{split}
\end{equation}
 We can express the inverse Fourier transform of $e^{2 \alpha(\mathbf{k})t}$ in terms of 
 modified Bessel functions of the first kind, indeed
 \begin{equation}
\begin{split}
   \left( \prod_{i=1}^d  \intB  \frac{a}{2\pi} dk_i \right)   e^{-i\sum_{j=1}^d {n_j k_j a}+ \alpha(\mathbf{k})t}
 =&  e^{-2(r+\tauInv) t}  \left(\prod_{i=1}^d  \intB  \frac{a}{2\pi} dk_i  \right)  
    e^{-i\sum_{j=1}^d n_j k_j a} e^{\frac{2t}{\tau d} \sum_{j=1}^d \cos( k_j)}\\
=&  e^{-2(r+\tauInv) t}  \prod_{i=1}^d  \left(\intB  \frac{a}{2\pi} dk_i    
    e^{-i  n_i k_i a} e^{\frac{2t}{\tau d} \cos( k_i)}\right)\\
=&  e^{-2(r+\tauInv) t}  \prod_{i=1}^d  \left(\intB  \frac{a}{2\pi} dk_i    
    e^{-i  n_i k_i a}  \sum_{m\in\mathbb{Z}} I_{m}( 2(\tau d)^{-1} t ) e^{i m k_i} \right)\\
=& e^{-2(r+\tauInv) t}  \sum_{m\in\mathbb{Z}} I_{m}( 2(\tau d)^{-1} t ) \delta_{n_im}\\
=& e^{-2(r+\tauInv) t}  \prod_{i=1}^d   I_{n_i}( 2(\tau d)^{-1} t ),\\
\end{split} 
\end{equation}
 where in the third step we used the generating function of modified Bessel functions defined in Eq. (\ref{BesselGener}). The two-point function in real space 
\begin{equation}\label{solGdcalReal}
\begin{split}
{\mathcal{G}} ( \mathbf{n}a,t) =& e^{-2(r+\tauInv ) t} \prod_{i=1}^d I_{n_i}( 2\taudInv t) +2r  
  \left( \prod_{i=1}^d  \intB  \frac{a}{2\pi} dk_i \right)  \int_0^t du e^{2\alpha(\mathbf{k})(t-u) -i\sum_i {n_i k_i a}} \hat{\mathcal{H}}(\mathbf{k},u)\\
 &+ \int_0^t du J(u)  e^{-2(r+\tauInv ) (t-u)} \prod_{i=1}^d I_{n_i}( 2\taudInv (t - u)) .\\
\end{split}
\end{equation}

We have not yet imposed the constraint   $\mathcal{G}(\mathbf{0},t)=1$, or calculated the source $J$. Let us do so in Laplace space.
 The Laplace transform of the constraint is ${\tilde{\mathcal{G}}}(\mathbf{0},s)=s^{-1}$. 
 On the other hand, we can express ${\tilde{\mathcal{G}}}(\mathbf{0},s)$ by  considering Eq. (\ref{solGdcalReal}) in the special case $\mathbf{n}=\mathbf{0}$, and taking the Laplace transform. 
 The convolution product is mapped to an ordinary product by the Laplace transform, hence
\begin{equation}\label{eqJ}
\frac{1}{s} =  \tilde{\varphi}_{0,d}(s) +2r \Psi_{0,d}(s) + \tilde{\varphi}_{0,d}(s)  \tilde{J}(s),
\end{equation}
with the notations
\begin{equation}\label{phindDef}
\begin{split}
 \varphi_{n,d}(t):=&  e^{-2(r+\tauInv ) t }  \left[ I_0( 2\taudInv t ) \right]^{d-1}( I_n( 2\taudInv t ),\\
 \Psi_{n,d}( s):=&  \left[ \prod_{i=1}^d  \left( \frac{a}{2\pi}\right) dk_i \right]
   e^{-ink_1a} \frac{1}{s - 2\alpha(\mathbf{k})}  \tilde{\hat{\mathcal{H}}}(k,s),\;\;\;\;\;\;(n\in\mathbb{Z}).\\
\end{split}
\end{equation}
 These notations generalize the ones introduced in Eq. (\ref{varPhiDef}) for the one-dimensional model, with 
$\varphi_{n,d} = \varphi_n$ and $\Psi_{n,d}=\Psi_n$.\\

The system is translationally invariant, moreover it is isotropic, hence (by applying the reasoning of Eq. (\ref{rhoDef}) in dimension $d$) we can express the density of domain walls at time $t$, denoted again by 
 $\varrho(t)$, in terms of the function $ {\mathcal{G}}$. Indeed, for any vertex $\mathbf{x}$ in the hypercubic lattice,
\begin{equation}
\begin{split}
 1-2\varrho(t)   =&   {\mathrm{Prob}}(s(\mathbf{x},t) = +1, s(\mathbf{x}+\mathbf{e_1},t) = +1) + {\mathrm{Prob}}(s(\mathbf{x},t) = -1, s(\mathbf{x}+\mathbf{e_1},t) = -1)\\
  &- {\mathrm{Prob}}(s(\mathbf{x},t) = -1, s(\mathbf{x}+\mathbf{e_1},t) = +1) - {\mathrm{Prob}}(s(\mathbf{x},t) = -1, s(\mathbf{x}+\mathbf{e_1},t) = +1) \\
=&\langle s(\mathbf{x},t) s(\mathbf{x}+\mathbf{e_1},t) \rangle \\
           =&  \mathcal{G}( a \mathbf{e_1}, t). 
\end{split}
\end{equation}
 The above equation generalizes Eq. (\ref{varrhoAndGcal}) to dimension $d$.\\

 Using Eq. (\ref{solGdcalReal}) in the special case $\mathbf{n} = \mathbf{e_1}$, the Laplace transform of the above equation is obtained as
\begin{equation}\label{sysd}
\frac{1}{s}-2\tilde{\varrho}(s) =  \tilde{\mathcal{G}}( \mathbf{e_1},s) =\tilde{\varphi}_{1,d}(s) +2r \Psi_{1,d}(s) + \tilde{\varphi}_{1,d}(s)  \tilde{J}(s).
\end{equation}

 Combining Eqs (\ref{eqJ},\ref{sysd}) yields a linear system in 
$\tilde{J}(s), \tilde{\varrho}(s)$. The solution reads
\begin{equation}
\begin{split}
\tilde{J}(s) &= \frac{1}{s\tilde{\varphi}_{0,d}(s)} - 2r \frac{\Psi_{0,d}(s)}{\tilde{\varphi}_{0,d}(s)} -1,\\
\tilde{\varrho}(s) &= \frac{1}{2}\left[  \frac{1}{s} - 2r \Psi_{1,d}(s) + 2r  \frac{\tilde{\varphi}_{1,d}(s)}{\tilde{\varphi}_{0,d}(s)}\Psi_{0,d}(s) - \frac{\tilde{\varphi}_{1,d}(s)}{s\tilde{\varphi}_{0,d}(s)} \right].
\end{split}
\end{equation}

  We obtain  the steady-state density of domain walls in dimension $d$ (denoted again by $\varrho_\infty$, as in Eq. (\ref{spot})) by applying the 
  final-value theorem:
\begin{equation}\label{spotd}
\begin{split}
\varrho_\infty =&  \underset{s\to 0}{\lim} ( s\tilde{\varrho}(s)) \\
=& \frac{1}{2}\underset{s\to 0}{\lim}\left(  1 - 2r s\Psi_{1,d}(s) + 2rs  \frac{\tilde{\varphi}_{1,d}(s)}{\tilde{\varphi}_{0,d}(s)}\Psi_{0,d}(s) - \frac{\tilde{\varphi}_{1,d}(s)}{\tilde{\varphi}_{0,d}(s)} \right).
\end{split} 
\end{equation}

 To obtain a more explicit expression of this density, we need  the Fourier--Laplace transform 
$\tilde{\hat{\mathcal{H}}}(k,s)$, which is the Laplace transform of Eq. (\ref{soldH}).
 The Laplace transforms  ${\tilde{\varphi}}_{0,d}(s)$ and
 ${\tilde{\varphi}}_{1,d}(s)$ are worked out in the appendix (Eq. (\ref{apo})) in integral form.
 The relevant limits are derived in the Appendix \ref{closedFormDerivation}. They result in the 
 following expression, which is the solution of Problem (II$_{(r)}$) in dimension $d$:
\begin{equation}\label{rhoExprFinal}
 \varrho_\infty = \frac{1}{4\mathcal{I}_d(2r\tau)} -\frac{r\tau}{2} 
- (r\tau)^2\left[   \mathcal{I}_d(2r\tau)
  +\frac{\mathcal{I}'_d(2r\tau)}{\mathcal{I}_d(2r\tau)} \right],
\end{equation}
where $\mathcal{I}_d(\epsilon)$ is the function  defined in Eq. (\ref{Id}).\\

%
%

\begin{figure}[H]
\begin{center}
\includegraphics[width=16cm]{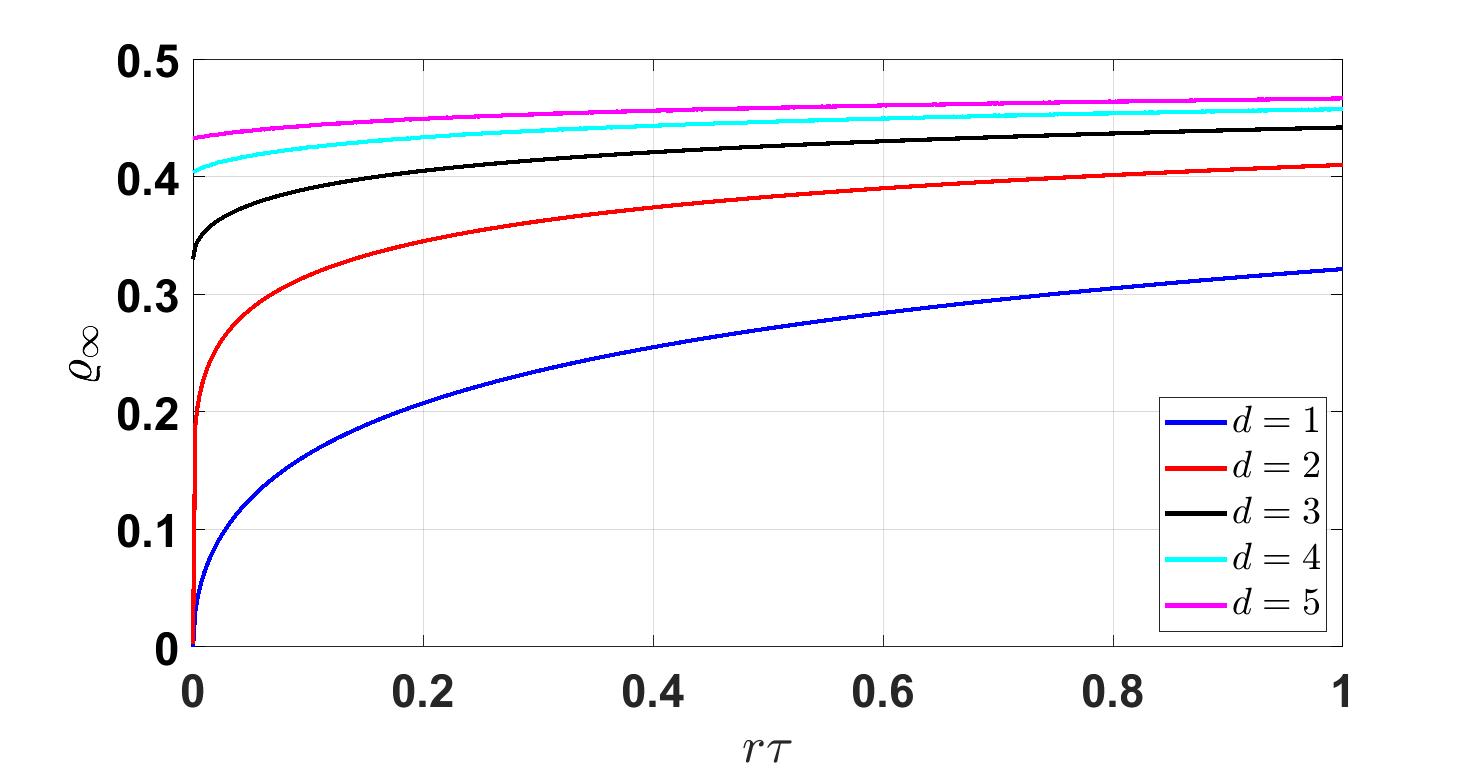} 
\caption{{\bf{The steady-state density of domain walls for initial conditions consisting of independent initially undecided voters.}} There is consensus in the limit of zero resetting rate in dimension $d\leq 2$. The tangent to the graph at $r=0$ is vertical for $d\leq 4$ (with a square-root singularity if $d=1$ or $d=3$, and an inverse-logarithmic singularity if $d=3$). It has a finite slope for $d\geq 5$, given by the susceptibility to the resetting rate expressed in Eq. (\ref{susceptibility}). }
 \label{figDensityHighDim}
\end{center}
\end{figure}

The steady-state density of domain walls is plotted on Fig. \ref{figDensityHighDim}
 for $d\leq 5$. It is an increasing function of the resetting rate, which is intuitive because 
 the resetting processes drive the system  away from consensus.\\

 The local behavior of ${\mathcal{I}}_d(\epsilon)$  for $\epsilon$ close to zero
 is worked out in Appendix \ref{local} (Eqs \ref{eqd2},\ref{eqd3},\ref{eqd4},\ref{eqd5}).
  These results complement the square-root singularity obtained in Eq. (\ref{equivRho}) in dimension one. They yield
 the behavior of $\varrho_\infty$ in the limit where the resetting rate $r$ is small compared 
 to the flipping rate $\tau^{-1}$: 
\begin{equation}
\varrho_\infty =
\begin{cases}
\sqrt{\frac{r\tau}{2}} + o\left(\sqrt{r\tau}   \right),\;\;\;\;\;\;\;\;\;&(d=1),\\
-\frac{\pi}{2\ln(r\tau)} + o\left( \frac{1}{\ln(r\tau)}\right) ,\;\;\;\;\;\;\;\;\;&(d=2),\\
\frac{1}{4 \mathcal{I}_3(0)} + \frac{6^{\frac{3}{2}}}{32\pi\mathcal{I}_3(0)^2}\sqrt{r\tau}  +   o( \sqrt{r\tau}),\;\;\;\;\;\;\;\;\;\;&(d=3),\\
 \frac{1}{4 \mathcal{I}_4(0)} - \frac{\pi^2}{ \mathcal{I}_4(0)^2}r\tau\ln( r\tau)+   o\left( r\tau \ln( r\tau) \right), \;\;\;\;\;\;\;\;\;&(d=4),\\
\frac{1}{4\mathcal{I}_{d}(0)} -\frac{1}{2}\left[1+
\frac{ \mathcal{I}'_{d}(0)}{\mathcal{I}_{d}(0)^2} \right]r\tau  + o( r\tau),\;\;\;\;\;\;\;&(d\geq 5).
\end{cases}
\end{equation} 
 In particular, $\varrho_\infty$ goes to zero when the resetting rate goes to zero if and only if $d\leq2$. This is consistent with the fact that he value of  $\varrho_{\infty}$ in the ordinary voter model  model is zero (meaning there is consensus in the steady state)  if and only if $d\leq 2$. Moreover,
 the non-zero  limit of $\varrho_\infty$ when $r$ goes to zero (in dimension  $d\geq 3$) is the 
 value of the steady-state density of domain walls in the ordinary voter model  \cite{frachebourg1996exact}.\\  


Moreover, if the dimension in greater than or equal to five, the steady-state density of domain walls has a finite susceptibility to the resetting rate:
\begin{equation}\label{susceptibility}
\left.\frac{\delta \varrho_\infty}{\delta r}\right|_{r=0} = -\frac{\tau}{2} \left(1+
\frac{ \mathcal{I}'_{d}(0)}{\mathcal{I}_{d}(0)^2} \right),\;\;\;\;\;\;\;\;\;\;(d\geq 5). 
\end{equation}


\section{Discussion}

 In this work we have introduced a toy model of stubborn voters on a hypercubic lattice by  resetting the opinion of each of the voters to its initial state at Poisson-distributed times. 
  The extra terms in the flipping rate corresponding to our local resetting prescription induce kinetic  equations for the one- and two-point functions. Our derivations are not based on a renewal argument (unlike models of single-particle systems under resetting considered for instance in  \cite{evans2018run,maso2022conditioned}).\\

 As in the ordinary voter model, the one-point function can be calculated exactly in any dimension. If the initial condition consists of a single decided voter at the origin, the average opinion state has the same 
 expression as the probability of presence of a diffusive  random walker on the lattice, whose position is stochastically reset to the origin. Moreover, for more general initial conditions, the one-point function is the convolution of this particular solution and the initial opinion-state profile.\\

 The evolution equation of the two-point function 
$G( \mathbf{x},\mathbf{y},t)=\langle s(\mathbf{x},t)s(\mathbf{y},t)\rangle$  contains terms involving the two-time two-point 
  function $H( \mathbf{x},\mathbf{y},t)=\langle s(\mathbf{x},t)s(\mathbf{y},0)\rangle$. This correlator is easily  calculated: for any fixed $\mathbf{y}$, it satisfies the same equation as the average opinion state.
Introducing a source at the origin to ensure that $G(\mathbf{x},\mathbf{x},t)=1$, we solved the evolution equation of  the two-point function in Laplace space, for initial conditions 
 consisting of independent undecided voters. The steady-state density of domain walls
 $\varrho_\infty$ follows in closed form in terms of Watson-like integrals. It is  a smooth, increasing  function of the quantity $r\tau$ on $]0,\infty[$. In dimension $d\geq 2$,  the density $\varrho_\infty$ of domain walls in the steady state goes to zero when the resetting rate $r$ becomes small 
  (compared to the maximum frequency $\tau^{-1}$ at which opinion states flip in the ordinary voter model). This limit is consistent with the consensus achieved in the ordinary model. There is a square-root  singularity at zero resetting rate in dimension one, and an inverse logarithmic singularity at zero resetting rate in dimension two.  For $d\geq 5$, the function is differentiable at $0$.\\ 

 We may ask how these results  could be applied to model real-world systems.
  The solution of the  ordinary voter model was motivated by the kinetics of the dimer-dimer
  reaction model \cite{krapivsky1992kinetics,evans1993kinetics,frachebourg1996exact}. 
 In this model, the sites of the lattice are filled with adsorbed monomers of two kinds, say $A$ and $B$.  Monomers adsorbed on two adjacent sites can react if they are of different kinds. The resulting dimer $AB$  immediately desorbs and is replaced  with an $A_2$ or $B_2$ dimer with equal probability, which reproduces the dynamics of the voter model (with $A$ and $B$ representing the two possible values of the opinion state). 
 In the dimer-dimer reaction model,  the resetting prescription is interpreted as follows: the substrate acts like a source of monomers, which replaces (at rate $r$) any adsorbed monomer with a monomer of the kind that was adsorbed at the same site in the initial configuration. The dimensions $d=1$ and $d=2$ are of particular interest, as they are the possible dimensions of a substrate: it is difficult to imagine 
 a realization of the dimer-dimer reaction model under resetting in dimension $d\geq 3$. The voter model under stochastic resetting is therefore more promising as a model of stylized facts in opinion formation in a social network, where connectivity can be high. Keeping the connectivity fixed throughout the network, one could try to make the model more realistic by allowing the resetting rate to vary across the lattice (for developments on  space-dependent resetting rate for Browian particles,  see \cite{roldan2017path,pinsky2020diffusive}). Moreover, it would be interesting to adapt the resetting prescription to heterogeneous networks, with uncorrelated  distributions of degrees as studied in  \cite{sood2005voter,sood2008voter}.\\

 Our resetting prescription is local in the sense that the opinion state at different sites are reset at independent times. Local resetting
may put the system in  a configuration that has never been explored before, on the occasion of the resetting of the opinion state of one voter.    Moreover, resetting the opinion state of a single voter does not decrease the number of voters that have never changed their opinion state since the beginning of the process. 
 It would  be interesting to explore the decay of the number of persistent spins \cite{derrida1995exact,derrida1996exact,ben1996coarsening} in the voter model under stochastic resetting.\\

\begin{appendices}
\section{Laplace transforms}\label{Laplace}
 
\subsection{Expressions needed in dimension one}\label{AppLap1d}
  To evaluate the Laplace transform of the modified Bessel functions of the first kind, we 
 used the identity
\begin{equation}\label{Itilden}
 \tilde{I}_n(s) = \frac{1}{\sqrt{s^2 - 1}}( s + \sqrt{s^2 - 1} )^{-|n|},
\end{equation}
which is the tabulated formula [4.16(1)] in \cite{abramowitz1988handbook}. It is valid for 
   $s>1$, and implies the expression of $\tilde{I}_n(1+r\tau)$  as reported  in Eq. (\ref{Kinf}). \\

  The Laplace transforms of the function denoted by $\varphi_n$ 
 in Eq. (\ref{varPhiDef}) is calculated by changing the integration variable to $u:=2\tauInv t$
 and using Eq. (\ref{Itilden}):
\begin{equation}\label{varPhiLap}
\begin{split}
 \tilde{\varphi_n}(s) =& \int_0^\infty dt  e^{-st} e^{-2(r+\tauInv)t}  I_n( 2\tau^{-1} t)\\
  =& \frac{1}{2\tau^{-1}} \int_0^\infty du   e^{-\frac{ s\tau }{2} u} e^{-(r\tau+1)  u} I_n(  u )\\
=& \frac{\tau}{2}   \tilde{I}_n\left(  r\tau + 1  + \frac{ s\tau }{2}\right)\\
=& \frac{\tau}{2}  \frac{1}{\sqrt{\left(r\tau + 1 + \frac{ s\tau }{2}\right)^2 - 1}}\left( r\tau + 1 + \frac{ s\tau }{2} + \sqrt{\left(r\tau+1+ \frac{ s\tau }{2}\right)^2 - 1} \right)^{-|n|},\;\;\;\;\;\;\;\;(n\in \mathbb{N}, s\geq 0).
\end{split}
\end{equation}
For a fixed value $r>0$ of the resetting rate, the function  $\tilde{\varphi_n}(s)$ has a finite limit when $s$ goes to zero. Hence the limit
\begin{equation}\label{limphi10}
\underset{s\to 0}{\lim}\frac{\tilde{\varphi_1}(s)}{\tilde{\varphi_0}(s)} =
\frac{1}{ r\tau+1 +\sqrt{(r\tau+1)^2 - 1 }},
\end{equation}
which is used in Eq. (\ref{rhoExpr}).\\

The quantity  $\Psi_n(s)$ introduced in Eq. (\ref{varPhiDef}) is expressed using the Fourier--Laplace transform of the two-point function $\mathcal{H}$, which is obtained from Eq. (\ref{Hkt}) by calculating integrals of exponential functions:
\begin{equation}
\begin{split}
\tilde{\hat{\mathcal{H}}}(k,s) =& \int_0^\infty dt \hat{\mathcal{H}}(k,s) e^{-st}  \\
=& \int_0^\infty dt  e^{-st}\left[ \exp \left( [\tau^{-1}\cos(ka) - (r+\tau^{-1}) ]t \right)  + \frac{r\left[  \exp \left( [\tau^{-1}\cos(ka) - (r+\tau^{-1}) ]t\right)-1\right]}{\tau^{-1}\cos(ka) - (r+\tau^{-1})}\right] \\
=& \frac{1 }{ s + r +\tauInv -\tauInv \cos( ka)}  \\
 &+\frac{r}{ - r -\tauInv +\tauInv \cos( ka)}\left( \frac{1 }{ s + r +\tauInv -\tauInv \cos( ka)}   -\frac{1}{s}\right).
\end{split}
\end{equation}
\begin{equation}
\begin{split}
\Psi_n(s) = \frac{a}{2\pi}& \intB dk \frac{e^{-inka}}{s + 2(r +\tauInv -\tauInv \cos( ka))} \tilde{\hat{H}}(k,s)\\
= \frac{1}{2\pi}&\int_{-\pi}^{\pi} dq \frac{ \cos( nq) }{s + 2(r +\tauInv -\tauInv \cos( q))}\left[
\frac{1 }{ s + r +\tauInv -\tauInv \cos( q)}  \right.\\
 &+\left.\frac{r}{ - r -\tauInv +\tauInv \cos( q)}\left( \frac{1 }{ s + r +\tauInv -\tauInv \cos( q)}   -\frac{1}{s}\right)\right],\;\;\;\;\;\;\;(n\in\mathbb{N}, s\geq 0),
\end{split}
\end{equation}
 where we have changed the integration variable to $q:= ka$ and used parity, since $e^{-inka}$ is multiplied in the integrand by 
 an even function of $k$ (a quantity that depends on $k$ through $\cos(ka)$ only). The last term in the above equation
 allows to read off the limit
\begin{equation}\label{limPsin}
\underset{s\to 0}{\lim}( rs\Psi_n(s)) = \frac{r^2\tau^2}{2\pi}\int_{-\pi}^{\pi} dx \frac{ \cos( nx) }{2( r\tau + 1 - \cos( x))^2}.
\end{equation}
This limit is used in the cases $n=0$ and $n=1$ in Eq. (\ref{rhoExpr}).\\

\subsection{Expressions needed in dimension $d$}

 The Laplace transform of the function $\varphi_{n,d}$ defined in Eq. (\ref{phindDef}) is 
 obtained in integral form by changing variable to $u = 2 \taudInv t$ and inserting the Fourier representation of each of the modified Bessel functions: 
\begin{equation}\label{apo}
\begin{split}
\tilde{\varphi}_{n,d}(s) =& \int_0^\infty dt e^{-(s+ 2r +2\tauInv)t} \left[ I_0( 2\taudInv t ) \right]^{d-1}( I_n( 2\taudInv t )\\
=&\frac{\tau d}{2} \int_0^\infty du e^{-\left( \frac{s\tau d}{2} + r\tau d + d \right)u}\left[ I_0( u) \right]^{d-1} I_n( u )\\
 =&\frac{\tau d}{2} \int_0^\infty du e^{-\left( \frac{s\tau d}{2} + r\tau d + d \right)u} \left[ \prod_{i=1}^d  \left( \frac{a}{2\pi}\right) \intB dk_i \right] e^{ink_d a} e^{u\sum_{i=1}^d \cos(k_i a )}\\
=&  \left[\prod_{i=1}^d  \left( \frac{a}{2\pi}\right) \intB dk_i \right] e^{ink_d a} \frac{\tau d}{2}\int_0^\infty du
 \exp\left(- u\left(  \frac{s\tau d}{2} + r\tau d + d  - \sum_{i=1}^d \cos(k_i a )\right)      \right)\\
=&  \left[\prod_{i=1}^d  \left( \frac{a}{2\pi}\right) \intB dk_i \right] e^{ink_d a} \int_0^\infty dv
 \exp\left(- v\left[  s + 2r + \frac{2}{\tau d} \left( d  - \sum_{i=1}^d \cos(k_i a ) \right)\right]      \right)\\
=& \left[\prod_{i=1}^d  \left( \frac{a}{2\pi}\right) \intB dk_i \right] e^{ink_d a}\frac{1}{s + 2r + \frac{2}{\tau d} \left( d  - \sum_{i=1}^d \cos(k_i a ) \right) }\\
=& \left[\prod_{i=1}^d  \int_{-\pi}^\pi \frac{dq_i}{2\pi} \right] \frac{\cos( nq_d)}{s + 2r + \frac{2}{\tau d} \left( d  - \sum_{i=1}^d \cos(q_i ) \right) }.
\end{split}
\end{equation}
In the last step we introduced the variables $q_i = a k_i$ for $i$ in $[1..d]$ and used parity.  The above expression is  used in Eq. (\ref{usedIn}).\\

Moreover, we need to take the Laplace transform of the quantity $\mathcal{H}(\mathbf{k},t)$ expressed in 
 Eq. (\ref{soldH})),
 to calculate the quantity $\Psi_{n,d}(s)$ defined in Eq. (\ref{phindDef}). 
  The operation  yields
\begin{equation}\label{PsindExpl} 
\Psi_{n,d}(s) = \left[ \prod_{i=1}^d  \int_{-\frac{\pi}{a}}^{+\frac{\pi}{a}}\left( \frac{a}{2\pi}\right) dk_i \right]
   e^{-ink_1a} \frac{1}{s - 2\alpha(\mathbf{k})}
\left[  \frac{1}{s-\alpha(\mathbf{k})}+\frac{r}{\alpha(\mathbf{k})}\left(\frac{1}{s - 2\alpha(\mathbf{k})} -\frac{1}{s}\right)  \right].
\end{equation}

\section{Expression of the steady-state density of domain walls in dimension $d$}\label{closedFormDerivation}

Let us work out the limits needed to express the steady-state density of domain 
 walls obtained in Eq. (\ref{spotd}).  
 The last term in Eq. (\ref{PsindExpl}), which is proportional to $s^{-1}$, yields the limit
\begin{equation}
\begin{split}
 \underset{s\to 0}{\lim}( 2rs \Psi_{n,d}(s)) =& 2r^2\left[ \prod_{i=1}^d  \int_{-\frac{\pi}{a}}^{+\frac{\pi}{a}}\left( \frac{a}{2\pi}\right) dk_i \right] \frac{e^{-ink_1a}}{2\alpha(\mathbf{k})^2}\\ 
=& (r\tau)^2\left[\prod_{i=1}^d  \int_{-\pi}^\pi \frac{dq_i}{2\pi} \right]   \frac{ \cos( nq_1)}{\left( r\tau +  \frac{1}{d}\sum_{i=1}^d (1 -\cos( q_i) \right)^2},\\
\end{split}
\end{equation}
 where we introduced the integration variables $q_i := ak_i, (1\leq i\leq d)$, and used parity.\\

Both of the expressions $\tilde{\varphi}_{0,d}(s)$  and $\tilde{\varphi}_{1,d}(s)$ have a finite limit when $s$ goes to zero because the resetting rate is positive. The cases $n=0$ and $n=1$ in Eq. (\ref{apo}) yield\\
\begin{equation}\label{usedIn}
\begin{split}
\underset{s\to 0}{\lim}\tilde{\varphi}_{0,d}(s) &=\tau {\mathcal{I}}_{d}(2r\tau),\\
 \underset{s\to 0}{\lim}\tilde{\varphi}_{1,d}(s) &=\tau {\mathcal{C}}_{d}(2r\tau),\\
\end{split}
\end{equation}
 where the function $\mathcal{I}_d$ is the one defined in Eq. (\ref{Id}), and
\begin{equation}\label{IdDef}
\begin{split}
{\mathcal{C}}_{d}(\epsilon)&:= \left[\prod_{i=1}^d  \int_{-\pi}^\pi \frac{dq_i}{2\pi} \right] \frac{\cos( q_d)}{ \epsilon + \frac{2}{ d}  \sum_{i=1}^d \left( 1- \cos(q_i ) \right) }.
\end{split}
\end{equation}

The density of domain walls in the steady state in dimension $d$ follows from 
 Eq. (\ref{spotd}) as
 \begin{equation}\label{rhoExprCI}
\begin{split}
  \varrho_\infty = 
\frac{1}{2}&\left[   1 -    (r\tau)^2\left(\prod_{i=1}^d  \int_{-\pi}^\pi \frac{dq_i}{2\pi} \right)   \frac{ \cos( q_1)}{\left( r\tau +  \frac{1}{d}\sum_{i=1}^d (1 -\cos( q_i) \right)^2}  \right.\\
& \left. - \frac{\mathcal{C}_{d}(2r\tau)}{ \mathcal{I}_{d}( 2r\tau )}\left( 1-  
(r\tau)^2\left[\prod_{i=1}^d  \int_{-\pi}^\pi \frac{dq_i}{2\pi} \right]   \frac{ 1}{\left( r\tau +  \frac{1}{d}\sum_{i=1}^d (1 -\cos( q_i) \right)^2}
\right) \right].\\
\end{split}
\end{equation}

The remaining integral terms in the above expression are related to the derivatives of the functions $\mathcal{I}_d$
 and $\mathcal{C}_d$ defined in Eqs (\ref{Id},\ref{IdDef}):
\begin{equation}\label{derIC}
\begin{split}
{\mathcal{I}}'_{d}(\epsilon)&=- \left[\prod_{i=1}^d  \int_{-\pi}^\pi \frac{dq_i}{2\pi} \right] \frac{1}{ \left[\epsilon + \frac{2}{ d}  \sum_{i=1}^d \left( 1 - \cos(q_i ) \right) \right]^2},\\
{\mathcal{C}}'_{d}(\epsilon)&=- \left[\prod_{i=1}^d  \int_{-\pi}^\pi \frac{dq_i}{2\pi} \right] \frac{\cos( q_d)}{ \left[\epsilon + \frac{2}{ d}  \sum_{i=1}^d \left( 1 - \cos(q_i ) \right) \right]^2 }.
\end{split}
\end{equation}

Substituting into Eq. (\ref{rhoExprCI}) yields
\begin{equation}\label{rhoExprCIDer}
\varrho_\infty = \frac{1}{2}\left[   1 + 4(r\tau)^2 {\mathcal{C}}'_{d}(2r\tau) 
   -\frac{\mathcal{C}_{d}(2r\tau)}{\mathcal{I}_{d}(2r\tau)}
 \left( 1 + 4(r\tau)^2  {\mathcal{I}}'_{d}(2r\tau) \right) \right].
\end{equation}

Moreover, we can easily express $\mathcal{C}_d(\epsilon)$ in terms of $\mathcal{I}_d(\epsilon)$, because the denominator in the integrand is symmetric in the variables $q_1,\dots,q_d$. Indeed, for any positive $\epsilon$,
\begin{equation}
\begin{split}
  1 =& \left( \prod_{i=1}^d \int_{-\pi}^\pi \frac{dq_i}{2\pi}\right)\frac{\epsilon + 2 -\frac{2}{d} \sum_{j=1}^d  \cos(q_j)}{\epsilon + 2 -\frac{2}{d} \sum_{j=1}^d \cos(q_j)}\\
=&( \epsilon + 2 ) \mathcal{I}_d(\epsilon) -d\times \frac{2}{d} \mathcal{C}_d(\epsilon),\\
\end{split}
\end{equation}
hence
\begin{equation}\label{eqCI}
\begin{split}
  \mathcal{C}_d(\epsilon) &= \left(\frac{\epsilon}{2} + 1\right) \mathcal{I}_d(\epsilon) -\frac{1}{2}.\\
\end{split}
\end{equation}
Differentiating Eq. (\ref{eqCI}) yields
\begin{equation}\label{eqCIprime}
\begin{split}
  \mathcal{C'}_d(\epsilon) &= \left(\frac{\epsilon}{2} + 1\right)  \mathcal{I'}_d(\epsilon) + \frac{1}{2}\mathcal{I}_d(\epsilon).
\end{split}
\end{equation}
Substituting Eqs (\ref{eqCI},\ref{eqCIprime}) into Eq. (\ref{rhoExprCIDer}) yields
 Eq. (\ref{rhoExprFinal}) reported in the main text.

%

\section{Local behaviour of the density of domain walls at small resetting rate}\label{local}

 The quantity $\mathcal{I}_d(\epsilon)$ is defined by an integral in Eq. (\ref{Id}).
 If $\epsilon=0$, the denominator of the integrand 
  takes the value $0$ at the origin (at $\mathbf{q}=\mathbf{0}$). 
 However, on a $(d-1)$-dimensional sphere of small radius $\rho$, the integrand is
 equivalent to $\rho^{-2}$, because
\begin{equation}
 \frac{1}{\frac{2}{d} \sum_{j=1}^d (1 - \cos(q_j))}
\underset{\rho\to 0,\\ \sum_{i=1}^d q_i^2 = \rho^2}\sim \frac{1}{ \frac{2}{d} \sum_{i=1}^d \frac{q_i^2}{2}} =  \frac{1}{ \rho^2}.
 \end{equation}
Moreover, the integration measure constributes a factor of $S_{d-1}\rho^{d-1} d\rho$ 
 (where $S_{d-1}$ denotes the measure of the $(d-1)$-dimensional unit sphere of equation $\sum_{i=1}^d q_i^2 = 1$).   The quantity $\mathcal{I}_d(\epsilon)$  therefore goes to infinity when $\epsilon$ goes to zero if $d\leq 2$ (and it goes to a finite limit denoted by $\mathcal{I}_d(0)$  if $d\geq 3$).\\

 On a sphere of radius $\rho$ centered at the origin,
the integrand in the expression of $\mathcal{I}'_d(\epsilon)$ (see Eq. (\ref{derIC})) is equivalent to 
$\rho^{-4}$ when the radius $\rho =||\mathbf{q}||$ goes to zero. Hence  ${\mathcal{I}}'_{d}(\epsilon)$ has a finite limit when $\epsilon$ goes to zero 
(and   $\mathcal{I}_d$ is differentiable at zero) if $d\geq 5$. Moreover, ${\mathcal{I}}'_{d}(\epsilon)$ goes to infinity when $\epsilon$ goes to zero if $d\leq 4$). These observations allow to work out equivalents of the quantity $\varrho_\infty$ at low resetting rate. \\

 In the following calculations, Taylor expansions and equivalents are meant in the limit  $r\tau\ll 1$. Repeated use is made of the change of integration variables from $q_i$ to 
 $y_i=  (2dr\tau)^{-\frac{1}{2}}q_i$, for $i$ in $[1..d]$. It is motivated  by the following Taylor expansion of the denominator, in the integrand involved in the definition of the function $\mathcal{I}_d(2r\tau)$:
\begin{equation}
\begin{split}
2r\tau + \frac{2}{d}\sum_{i=1}^d\left(1-\cos( q_i)\right) = 2r\tau + \frac{2}{d}\sum_{i=1}^d (2dr\tau)\frac{y_i^2}{2}+ o(r\tau)
\underset{r\tau \ll 1}{\sim}(2r\tau)\left( 1 + \sum_{i=1}^d y_i^2\right).
\end{split}
\end{equation}

\subsection{Dimension $d\geq 5$}\label{eqd5}
As the function $\mathcal{I}_{d}$ is differentiable at $0$, we obtain the following   Taylor expansion of Eq. (\ref{rhoExprFinal}) at order $1$ in $r\tau \ll 1$:
\begin{equation}\label{eqd5}
\begin{split}
\varrho_\infty =& \frac{1}{4\left[\mathcal{I}_d(0) + 2r\tau\mathcal{I}'_d(0) +o(r\tau) \right]} -\frac{r\tau}{2} + o(r\tau)\\
=&\frac{1}{4\mathcal{I}_d(0) } \left[1 - 2r\tau\frac{\mathcal{I}'_d(0)}{\mathcal{I}_d(0)} +o(r\tau) \right]-\frac{r\tau}{2} + o(r\tau)\\
=& \frac{1}{4\mathcal{I}_d(0) } -\frac{1}{2} \left[1 + \frac{\mathcal{I}'_d(0)}{\mathcal{I}_d(0)^2}  \right]r\tau + o(r\tau).
\end{split}
\end{equation}

\subsection{Dimension $d=4$}

The function $\mathcal{I}_4$ is not differentiable at $0$.  The change of variables
\begin{equation}
 y_i:=\frac{q_i}{\sqrt{8r\times \tau}},\;\;\;\;\;\;\;\;\;\;\;\;(1\leq i \leq d),
\end{equation}
 yields 
\begin{equation}\label{eqExp4}
\begin{split}
\mathcal{I}_4(2r\tau) - \mathcal{I}_4(0) &= \left( \prod_{i=1}^d \int_{-\pi}^\pi \frac{dq_i}{2\pi} \right)
 \frac{2r\tau}{ \left(2r\tau +  \frac{2}{d} \sum_{j=1}^d(1 - \cos(q_j) )\right) \left(  \frac{2}{d} \sum_{j=1}^d(1 - \cos(q_j) )\right) }\\
&\underset{r\tau \ll 1}{\sim} -\frac{2r\tau(2r\tau d)^{\frac{d}{2}}}{(2r\tau)^2}
 \left( \prod_{i=1}^4 \int_{-\frac{\pi}{\sqrt{8r\tau}}}^{\frac{\pi}{\sqrt{8r\tau}}} \frac{dy_i}{2\pi} \right)
 \frac{1}{\left( 1+ \sum_{j=1}^d y_j^2 \right) \left( \sum_{k=1}^d y_k^2 \right) }\\
 &\sim  -\frac{2r\tau(8r\tau)^2}{(2r\tau)^2} \frac{S_3}{(2\pi)^4} \int_0^\frac{\pi}{\sqrt{8r\tau}}\frac{\rho^3 d\rho}{(1+\rho^2)\rho^2} \\
&\sim   +\frac{4}{\pi^2}r\tau \ln( \sqrt{r\tau}) = \frac{2}{\pi^2}r\tau \ln( r\tau ),\\
\end{split}
\end{equation}
 where $S_3=2\pi^2$ is the area of the three-sphere. 
 The quantity $\mathcal{I}'_4(2r\tau)$ diverges logarithmically when $r\tau$ goes to zero, indeed
\begin{equation}
\begin{split}
{\mathcal{I}}'_{4}(2r\tau)=&- \left(\prod_{i=1}^d  \int_{-\pi}^\pi \frac{dq_i}{2\pi} \right) \frac{1}{ \left[2r\tau + \frac{2}{ d}  \sum_{i=1}^d \left( 1 - \cos(q_i ) \right) \right]^2} \\
\underset{r\tau \ll 1}{\sim}& \frac{(8r\tau)^{\frac{4}{2}}}{(2r\tau)^2}\frac{S_3}{(2\pi)^3} \int_0^{\frac{\pi}{\sqrt{8r\tau}}} \frac{\rho^3 d\rho}{(1+\rho^2)^2}
= O\left( \ln(r\tau)\right).
\end{split}
\end{equation}
 Moreover, $\mathcal{I}_4(0)$ is finite. Hence
 all the terms in the expression of $\varrho_\infty$ apart from $(4 \mathcal{I}_d(2r\tau))^{-1}$  are negligible compared to $r\tau \ln( r\tau)$:
\begin{equation} 
\begin{split}
(r\tau)^2 \frac{\mathcal{I}'_4( 2r\tau)}{\mathcal{I}_4(0)^2} &= o( r\tau \ln(r\tau)),\\
r\tau &= o( r\tau \ln(r\tau)),\\
(r\tau)^2 \mathcal{I}_4( 2r\tau) &= o(  r\tau \ln(r\tau)).\\
\end{split}
\end{equation}
 Combining with the equivalent of $\mathcal{I}_4(2r\tau) - \mathcal{I}_4(0)$ derived in Eq. (\ref{eqExp4}), we obtain the following expansion of the density of domain walls when $r\tau \ll 1$:
\begin{equation}\label{eqd4}
\begin{split}
\varrho_\infty =& \frac{1}{4\left[ \mathcal{I}_{4}(0) + \frac{2}{\pi^2} r\tau \ln(r\tau) +o\left( r\tau \ln( r\tau) \right) \right]} + o\left( r\tau \ln( r\tau) \right)\\
=& \frac{1}{4 \mathcal{I}_4(0)} - \frac{1}{2\pi^2 \mathcal{I}_4(0)^2}r\tau\ln( r\tau)+   o\left( r\tau \ln( r\tau) \right)\\
 \simeq& 0.4036 - 0.1320 r\tau\ln( r\tau)+   o\left( r\tau \ln( r\tau) \right).
\end{split}
\end{equation}

\subsection{Dimension $d=3$}

The function $\mathcal{I}_3$ is not differentiable at $0$. The change of variables
\begin{equation}
 y_i:=\frac{q_i}{\sqrt{6 \times r\tau}},\;\;\;\;\;\;\;\;(1\leq i \leq 3),
\end{equation}
 yields 
\begin{equation}\label{expa}
\begin{split}
\mathcal{I}_3(2r\tau) - \mathcal{I}_3(0) &=\left(\prod_{i=1}^3 \int_{-\pi}^\pi \frac{dq_i}{2\pi}\right)
 \frac{2r\tau}{ \left(2r\tau +  \frac{2}{3} \sum_{j=1}^3(1 - \cos(q_j) )\right) \left(  \frac{2}{3} \sum_{j=1}^3(1 - \cos(q_j) )\right) }\\
&\underset{r\tau \ll 1}{\sim} -\frac{2r\tau(6r\tau )^{\frac{3}{2}}}{(2r\tau)^2}
 \left( \prod_{i=1}^3 \int_{-\frac{\pi}{\sqrt{6r\tau}}}^{\frac{\pi}{\sqrt{6r\tau}}} \frac{dy_i}{2\pi} \right)
 \frac{1}{\left( 1+ \sum_{j=1}^3 y_j^2 \right) \left( \sum_{k=1}^3 y_k^2 \right)},\\
&\sim \frac{6^{3/2}}{2} \sqrt{r\tau} \times 4\pi \frac{1}{8\pi^3} \int_0^{\infty} \frac{d\rho}{1+\rho^2} =  -   \frac{6^{3/2}}{8\pi} \sqrt{r\tau}.\\
\end{split}
\end{equation}
The quantity $\mathcal{I}_3(0)$ is known explicitly as one of the  Watson integrals \cite{watson1939three,glasser1977extended,zucker201170+}:
\begin{equation}\label{Watson3}
 \mathcal{I}_3(0) = \frac{\sqrt{6}}{64\pi^3}\Gamma\left( \frac{1}{24}\right)
\Gamma\left( \frac{5}{24}\right)\Gamma\left( \frac{7}{24}\right)\Gamma\left( \frac{11}{24}\right)\simeq 0.7582.
\end{equation}
The same change of variables as above yields
\begin{equation}
\begin{split}
\mathcal{I}'_{3}(2r\tau) &= - \left(\prod_{i=1}^3  \int_{-\pi}^\pi \frac{dq_i}{2\pi} \right) \frac{1}{ \left[2r\tau + \frac{2}{ d}  \sum_{i=1}^d \left( 1 - \cos(q_i) \right) \right]^2}\\
&\underset{r\tau \ll 1}{\sim} \frac{(6r\tau)^{\frac{3}{2}}}{(2r\tau)^2}
 \left( \prod_{i=1}^3 \int_{-\frac{\pi}{\sqrt{6r\tau}}}^{\frac{\pi}{\sqrt{6r\tau}}} \frac{dy_i}{2\pi} \right)
\frac{1}{\left(1+\sum_{i=1}^3 y_i^2\right)^2}\\
&\sim \frac{(6r\tau)^{\frac{3}{2}}}{(2\pi)^3(2r\tau)^2}\times 4\pi\int_0^{\infty} \frac{\rho^2 d\rho}{(1+\rho^2)^2}\\
&= O\left( (r\tau)^{-\frac{1}{2}}\right).
\end{split}
\end{equation}

Hence the following terms in the expression of $\varrho_\infty$ in Eq. (\ref{rhoExprFinal}) are negligible compared to $\sqrt{r\tau}$ at low resetting rate:
\begin{equation}
\begin{split}
r\tau =& o( \sqrt{ r\tau} ),\\
(r\tau)^2\mathcal{I}_3(r\tau) &= O( (r\tau)^2) =o( \sqrt{ r\tau} ),\\
(r\tau)^2 \frac{\mathcal{I}'_3( 2r\tau)}{\mathcal{I}_3(0)} &=   O( (r\tau)^{\frac{3}{2}})  =  o(\sqrt{r\tau} ).
\end{split}
\end{equation}

Combining with the expansion in Eq. (\ref{expa}) yields
\begin{equation}\label{eqd3}
\begin{split}
\varrho_\infty =& \frac{1}{4} \frac{1}{\mathcal{I}_{3}(0) -     \times  \frac{6^{3/2}}{8\pi} \sqrt{r\tau}+ o(\sqrt{r\tau})}  + o\left(  \sqrt{ r\tau} \right)\\
=& \frac{1}{4 \mathcal{I}_3(0)} + \frac{6^{\frac{3}{2}}}{32\pi\mathcal{I}_3(0)^2}\sqrt{r\tau}  +   o( \sqrt{r\tau})\\
\simeq& 0.3297 + 0.25431 \sqrt{r\tau}+   o( \sqrt{r\tau}).
\end{split}
\end{equation}


\subsection{Dimension $d=2$}
The quantity $\mathcal{I}_{2}(2r\tau)$  goes logarithmically to infinity 
 when $r$ goes to $0$. Indeed,
 changing variables to $y_i = q_i/(\sqrt{4r\tau})$ (for $i$ in $\{1,2\}$) yields: 
\begin{equation}\label{I2Dev}
\begin{split}
\mathcal{I}_{2}(2r\tau) &= \int_{-\pi}^\pi \frac{dq_1}{2\pi}\int_{-\pi}^\pi \frac{dq_2}{2\pi}
\frac{1}{\left( 2r\tau + 2 - \cos( q_1) -\cos( q_2)\right)}\\
&\underset{r\tau \ll 1}{\sim}\frac{ 4r\tau }{2r\tau}\int_{-\pi/\sqrt{4r\tau}}^{\pi/\sqrt{4r\tau}} \frac{dy_1}{2\pi}\int_{-\pi/\sqrt{4r\tau}}^{\pi/\sqrt{4r\tau}} \frac{dy_2}{2\pi}
\frac{1}{1 + y_1^2 + y_2^2}\\
&\sim \frac{ 4r\tau }{(2\pi)^2 (2r\tau)} \times 2\pi\int_{0}^{\pi/\sqrt{4r\tau}} \frac{\rho d\rho}{(1+\rho^2)}\\
 &\sim -\frac{\ln( r\tau)}{2\pi},\\
\end{split}
\end{equation}
hence the equivalent
\begin{equation}
\frac{1}{4\mathcal{I}_2(2r\tau)} \underset{r\tau \ll 1}{\sim}  -\frac{\pi}{2\ln(r\tau)}.
\end{equation}

Moreover,
\begin{equation}
\begin{split}
\mathcal{I}'_{2}(2r\tau) &= \int_{-\pi}^\pi \frac{dq_1}{2\pi}\int_{-\pi}^\pi \frac{dq_2}{2\pi}
\frac{1}{\left( 2r\tau + 2 - \cos( q_1) -\cos( q_2)\right)^2}\\
&\underset{r\tau \ll 1}{\sim}\frac{ 4r\tau }{(2r\tau)^2}\int_{-\pi/\sqrt{4r\tau}}^{\pi/\sqrt{4r\tau}} \frac{dy_1}{2\pi}\int_{-\pi/\sqrt{4r\tau}}^{\pi/\sqrt{4r\tau}} \frac{dy_2}{2\pi}
\frac{1}{\left( 1 + y_1^2 + y_2^2)\right)^2}\\
&\sim \frac{ 4r\tau }{(2\pi)^2(2r\tau)^2} \times 2\pi\int_{0}^{\pi/\sqrt{4r\tau}} 
\frac{\rho d\rho}{(1+\rho^2)^2}\\
&= O( (r\tau)^{-1}).\\
\end{split}
\end{equation}

Hence the following terms in $\varrho_\infty$ are negligible compared to $o\left( (\ln(r\tau))^{-1}\right)$ at low resetting rate:
\begin{equation}
\begin{split}
r\tau =& o\left( (\ln(r\tau))^{-1}\right),\\
(r\tau)^2\mathcal{I}_2(r\tau) &= O(   (r\tau)^2\ln(r\tau)) = o( r\tau ) =o( \sqrt{ r\tau} ) = o\left( (\ln(r\tau))^{-1}\right),\\
(r\tau)^2 \frac{\mathcal{I}'_2( 2r\tau)}{\mathcal{I}_2( 2r\tau)} &=  O( r\tau(\ln(r\tau))^{-1} )
 = o\left( (\ln(r\tau))^{-1}\right).
\end{split}
\end{equation}
Hence the density $\varrho_\infty$ is equivalent to $(4\mathcal{I}_2(2r\tau))^{-1}$ at small resetting rate:
\begin{equation}\label{eqd2}
\varrho_\infty \underset{r\tau \ll 1}{\sim}  -\frac{\pi}{2\ln(r\tau)}.
\end{equation}

\subsection{Dimension $d=1$}\label{completeness}

Starting from the expansion 
 \begin{equation}
\begin{split}
\frac{1}{ r\tau+1 +\sqrt{(r\tau+1)^2 - 1 }} = \frac{1}{ r\tau+1 +\sqrt{r \tau} \sqrt{ 2 + r\tau}} = 1 - \sqrt{2r\tau} + o( r \tau),
\end{split}
\end{equation}
 we obtain the equivalent at low resetting rate of the limit worked out in Eq. (\ref{limphi10})
\begin{equation}\label{quandcons}
\begin{split}
1 - \frac{\tilde{\varphi}_1(0)}{\tilde{\varphi}_0(0)} =  \sqrt{2r\tau} + o( r\tau).
\end{split}
\end{equation}

For completeness, let us work out the same equivalent starting from the general expression 
 of  the steady-state density of domain walls in terms of the function $\mathcal{I}_1$ in Eq. (\ref{rhoExprFinal}). To ensure consistency with Eq. (\ref{rhoExpr}), the expression $(4\mathcal{I}_1(2r\tau))^{-1}$ should behave in the same way as the quantity $\frac{1}{2}\left(1 - \frac{\tilde{\varphi}_1(0)}{\tilde{\varphi}_0(0)}\right)$ in the regime $r\tau\ll 1$. 
The quantity $\mathcal{I}_{1}(\epsilon)$ goes to infinity 
 when $\epsilon$ goes to $0$.  Changing the integration variable to $y := x/(\sqrt{2r\tau})$ yields 
\begin{equation}\label{I1Dev}
\begin{split}
\mathcal{I}_{1}(2r\tau) =& \sqrt{2r\tau} \int_{-\frac{\pi}{\sqrt{2r\tau}}}^{\frac{\pi}{\sqrt{2r\tau}}}\frac{dy}{2\pi}  \frac{ 1 }{2r\tau( 1 +  y^2)( 1 + o(1)) }
\underset{r\ll\tau}{\sim} \frac{1}{2\sqrt{2r\tau}}.\\
\end{split}
\end{equation}
  Hence
\begin{equation}
\begin{split}
\frac{1}{4  \mathcal{I}_{1}(2r\tau)}
&\underset{r\tau \ll 1}{\sim}\frac{\sqrt{2r\tau}}{2},
\end{split}
\end{equation}
 which is consistent with Eq. (\ref{quandcons}).\\

 On the other hand, the integral terms in Eq. (\ref{rhoExpr}) diverge when the resetting rate goes to zero.
 We can work out an equivalent of these integral terms  as follows:
\begin{equation}\label{extra1}
\begin{split}
  \underset{s\to0}{\lim}  \left[- 2r s\Psi_1(s) + 2rs  \frac{\tilde{\varphi}_1(s)}{\tilde{\varphi}_0(s)}\Psi_0(s)\right] 
\underset{r\tau \ll 1}{\sim}    \left( \frac{(r\tau)^2}{2\pi} \int_{-\pi}^\pi  dx \frac{1-\cos( x )}{(1 - \cos( x ) + r\tau )^2}\right).
\end{split}
\end{equation}
\begin{equation}\label{extra2}
\begin{split}
 (r\tau)^2 \int_{-\pi}^\pi  dx \frac{1-\cos( x )}{(1 - \cos( x ) + r\tau )^2}
 =& (r\tau)^2 
 \int_{-\pi}^\pi  dx\frac{2\sin^2\left(  \frac{x}{2} \right)}{ \left( 2\sin^2\left(  \frac{x}{2}\right) + r\tau\right)^2}\\
=& (r\tau)^2\sqrt{2 r\tau} \int_{-\frac{\pi}{\sqrt{2 r\tau}}}^{\frac{\pi}{\sqrt{2 r\tau}}} dy
 \frac{2\sin^2 \left( \frac{y\sqrt{r\tau}}{\sqrt{2}}\right)}{\left( 2\sin^2\left(  \frac{y\sqrt{r\tau}}{\sqrt{2}}\right) +  r\tau\right)^2}\\
=& (r\tau)^2 \frac{\sqrt{2 r\tau}}{r\tau} \int_{-\infty}^{\infty} dy
 \frac{y^2}{ (y^2 + 1)^2}(1+o(1))\\
\underset{r\ll \tauInv}{=} & O (r\tau)^{\frac{3}{2}}.
\end{split}
\end{equation}
where we have changed the integration variable $y:=x/(\sqrt{2 r\tau})$. The contribution of the integral terms to the density 
 $\varrho_\infty$ at low resetting rate is therefore subdominant compared to $\sqrt{r\tau}$, which yields the equivalent displayed in Eq. (\ref{equivRho}).\\

\end{appendices}

\bibliography{bibRefsNewJun2023} 

\begin{thebibliography}{10}

\bibitem{krapivsky1992kinetics}
P.~L. Krapivsky, ``Kinetics of monomer-monomer surface catalytic reactions,''
  {\em Physical Review A}, vol.~45, no.~2, p.~1067, 1992.

\bibitem{frachebourg1996exact}
L.~Frachebourg and P.~L. Krapivsky, ``Exact results for kinetics of catalytic
  reactions,'' {\em Physical Review E}, vol.~53, no.~4, p.~R3009, 1996.

\bibitem{liggett1985interacting}
T.~M. Liggett and T.~M. Liggett, {\em Interacting particle systems}, vol.~2.
\newblock Springer, 1985.

\bibitem{redner2019reality}
S.~Redner, ``Reality-inspired voter models: A mini-review,'' {\em Comptes
  Rendus Physique}, vol.~20, no.~4, pp.~275--292, 2019.

\bibitem{kineticView}
P.~L. Krapivsky, S.~Redner, and E.~Ben-Naim, {\em A kinetic view of statistical
  physics}.
\newblock Cambridge University Press, 2010.

\bibitem{miron2021diffusion}
A.~Miron and S.~Reuveni, ``Diffusion with local resetting and exclusion,'' {\em
  Physical Review Research}, vol.~3, no.~1, p.~L012023, 2021.

\bibitem{evans2011diffusion}
M.~R. Evans and S.~N. Majumdar, ``Diffusion with stochastic resetting,'' {\em
  Physical review letters}, vol.~106, no.~16, p.~160601, 2011.

\bibitem{evans2011optimal}
M.~R. Evans and S.~N. Majumdar, ``Diffusion with optimal resetting,'' {\em
  Journal of Physics A: Mathematical and Theoretical}, vol.~44, no.~43,
  p.~435001, 2011.

\bibitem{pal2015diffusion}
A.~Pal, ``Diffusion in a potential landscape with stochastic resetting,'' {\em
  Physical Review E}, vol.~91, no.~1, p.~012113, 2015.

\bibitem{kusmierz2014first}
L.~Kusmierz, S.~N. Majumdar, S.~Sabhapandit, and G.~Schehr, ``First order
  transition for the optimal search time of l{\'e}vy flights with resetting,''
  {\em Physical review letters}, vol.~113, no.~22, p.~220602, 2014.

\bibitem{evans2018run}
M.~R. Evans and S.~N. Majumdar, ``Run and tumble particle under resetting: a
  renewal approach,'' {\em Journal of Physics A: Mathematical and Theoretical},
  vol.~51, no.~47, p.~475003, 2018.

\bibitem{refractory}
M.~R. Evans and S.~N. Majumdar, ``Effects of refractory period on stochastic
  resetting,'' {\em Journal of Physics A: Mathematical and Theoretical},
  vol.~52, no.~1, p.~01LT01, 2018.

\bibitem{kumar2020active}
V.~Kumar, O.~Sadekar, and U.~Basu, ``Active brownian motion in two dimensions
  under stochastic resetting,'' {\em Physical Review E}, vol.~102, no.~5,
  p.~052129, 2020.

\bibitem{topical}
M.~R. Evans, S.~N. Majumdar, and G.~Schehr, ``Stochastic resetting and
  applications,'' {\em Journal of Physics A: Mathematical and Theoretical},
  vol.~53, no.~19, p.~193001, 2020.

\bibitem{gupta2022stochastic}
S.~Gupta and A.~M. Jayannavar, ``Stochastic resetting: A (very) brief review,''
  {\em Frontiers in Physics}, vol.~10, p.~789097, 2022.

\bibitem{grange2021aggregation}
P.~Grange, ``Aggregation with constant kernel under stochastic resetting,''
  {\em Journal of Physics A: Mathematical and Theoretical}, 2021.

\bibitem{pelizzola2021simple}
A.~Pelizzola, M.~Pretti, and M.~Zamparo, ``Simple exclusion processes with
  local resetting,'' {\em Europhysics Letters}, vol.~133, no.~6, p.~60003,
  2021.

\bibitem{magoni2020ising}
M.~Magoni, S.~N. Majumdar, and G.~Schehr, ``Ising model with stochastic
  resetting,'' {\em Physical Review Research}, vol.~2, no.~3, p.~033182, 2020.

\bibitem{aron2020nonanalytic}
C.~Aron and M.~Kulkarni, ``Nonanalytic nonequilibrium field theory: Stochastic
  reheating of the ising model,'' {\em Physical Review Research}, vol.~2,
  no.~4, p.~043390, 2020.

\bibitem{basu2019symmetric}
U.~Basu, A.~Kundu, and A.~Pal, ``Symmetric exclusion process under stochastic
  resetting,'' {\em Physical Review E}, vol.~100, no.~3, p.~032136, 2019.

\bibitem{sadekar2020zero}
O.~Sadekar and U.~Basu, ``Zero-current nonequilibrium state in symmetric
  exclusion process with dichotomous stochastic resetting,'' {\em Journal of
  Statistical Mechanics: Theory and Experiment}, vol.~2020, no.~7, p.~073209,
  2020.

\bibitem{karthika2020totally}
S.~Karthika and A.~Nagar, ``Totally asymmetric simple exclusion process with
  resetting,'' {\em Journal of Physics A: Mathematical and Theoretical},
  vol.~53, no.~11, p.~115003, 2020.

\bibitem{quetzalcoatl2019predator}
J.~Quetzalcoatl Toledo-Marin, D.~Boyer, and F.~J. Sevilla, ``Predator-prey
  dynamics: Chasing by stochastic resetting,'' {\em arXiv e-prints},
  pp.~arXiv--1912, 2019.

\bibitem{evans2022exactly}
M.~R. Evans, S.~N. Majumdar, and G.~Schehr, ``An exactly solvable predator prey
  model with resetting,'' {\em Journal of Physics A: Mathematical and
  Theoretical}, vol.~55, no.~27, p.~274005, 2022.

\bibitem{mercado2018lotka}
G.~Mercado-V{\'a}squez and D.~Boyer, ``Lotka--{V}olterra systems with
  stochastic resetting,'' {\em Journal of Physics A: Mathematical and
  Theoretical}, vol.~51, no.~40, p.~405601, 2018.

\bibitem{gupta2014fluctuating}
S.~Gupta, S.~N. Majumdar, and G.~Schehr, ``Fluctuating interfaces subject to
  stochastic resetting,'' {\em Physical review letters}, vol.~112, no.~22,
  p.~220601, 2014.

\bibitem{gupta2016resetting}
S.~Gupta and A.~Nagar, ``Resetting of fluctuating interfaces at power-law
  times,'' {\em Journal of Physics A: Mathematical and Theoretical}, vol.~49,
  no.~44, p.~445001, 2016.

\bibitem{sarkar2022synchronization}
M.~Sarkar and S.~Gupta, ``Synchronization in the {K}uramoto model in presence
  of stochastic resetting,'' {\em Chaos: An Interdisciplinary Journal of
  Nonlinear Science}, vol.~32, no.~7, 2022.

\bibitem{durang2014statistical}
X.~Durang, M.~Henkel, and H.~Park, ``The statistical mechanics of the
  coagulation--diffusion process with a stochastic reset,'' {\em Journal of
  Physics A: Mathematical and Theoretical}, vol.~47, no.~4, p.~045002, 2014.

\bibitem{grange2020entropy}
P.~Grange, ``Entropy barriers and accelerated relaxation under resetting,''
  {\em Journal of Physics A: Mathematical and Theoretical}, 2020.

\bibitem{ZRPSS}
P.~Grange, ``Steady states in a non-conserving zero-range process with
  extensive rates as a model for the balance of selection and mutation,'' {\em
  Journal of Physics A: Mathematical and Theoretical}, vol.~52, no.~36,
  p.~365601, 2019.

\bibitem{ZRPResetting}
P.~Grange, ``Non-conserving zero-range processes with extensive rates under
  resetting,'' {\em Journal of Physics Communications}, vol.~4, no.~4,
  p.~045006, 2020.

\bibitem{nagar2023stochastic}
A.~Nagar and S.~Gupta, ``Stochastic resetting in interacting particle systems:
  A review,'' {\em Journal of Physics A: Mathematical and Theoretical}, 2023.

\bibitem{evans2014diffusion}
M.~R. Evans and S.~N. Majumdar, ``Diffusion with resetting in arbitrary spatial
  dimension,'' {\em Journal of Physics A: Mathematical and Theoretical},
  vol.~47, no.~28, p.~285001, 2014.

\bibitem{maso2022conditioned}
A.~Mas{\'o}-Puigdellosas, D.~Campos, and V.~M{\'e}ndez, ``Conditioned backward
  and forward times of diffusion with stochastic resetting: A renewal theory
  approach,'' {\em Physical Review E}, vol.~106, no.~3, p.~034126, 2022.

\bibitem{evans1993kinetics}
J.~W. Evans and T.~Ray, ``Kinetics of the monomer-monomer surface reaction
  model,'' {\em Physical Review E}, vol.~47, no.~2, p.~1018, 1993.

\bibitem{roldan2017path}
{\'E}.~Rold{\'a}n and S.~Gupta, ``Path-integral formalism for stochastic
  resetting: Exactly solved examples and shortcuts to confinement,'' {\em
  Physical Review E}, vol.~96, no.~2, p.~022130, 2017.

\bibitem{pinsky2020diffusive}
R.~G. Pinsky, ``Diffusive search with spatially dependent resetting,'' {\em
  Stochastic Processes and their Applications}, vol.~130, no.~5,
  pp.~2954--2973, 2020.

\bibitem{sood2005voter}
V.~Sood and S.~Redner, ``Voter model on heterogeneous graphs,'' {\em Physical
  review letters}, vol.~94, no.~17, p.~178701, 2005.

\bibitem{sood2008voter}
V.~Sood, T.~Antal, and S.~Redner, ``Voter models on heterogeneous networks,''
  {\em Physical Review E}, vol.~77, no.~4, p.~041121, 2008.

\bibitem{derrida1995exact}
B.~Derrida, V.~Hakim, and V.~Pasquier, ``Exact first-passage exponents of 1d
  domain growth: relation to a reaction-diffusion model,'' {\em Physical review
  letters}, vol.~75, no.~4, p.~751, 1995.

\bibitem{derrida1996exact}
B.~Derrida, V.~Hakim, and V.~Pasquier, ``Exact exponent for the number of
  persistent spins in the zero-temperature dynamics of the one-dimensional
  potts model,'' {\em Journal of statistical physics}, vol.~85, no.~5,
  pp.~763--797, 1996.

\bibitem{ben1996coarsening}
E.~Ben-Naim, L.~Frachebourg, and P.~L. Krapivsky, ``Coarsening and persistence
  in the voter model,'' {\em Physical Review E}, vol.~53, no.~4, p.~3078, 1996.

\bibitem{abramowitz1988handbook}
M.~Abramowitz, I.~A. Stegun, and R.~H. Romer, ``Handbook of mathematical
  functions with formulas, graphs, and mathematical tables,'' 1988.

\bibitem{watson1939three}
G.~N. Watson, ``Three triple integrals,'' {\em The Quarterly Journal of
  Mathematics}, no.~1, pp.~266--276, 1939.

\bibitem{glasser1977extended}
M.~L. Glasser and I.~J. Zucker, ``Extended {W}atson integrals for the cubic
  lattices,'' {\em Proceedings of the National Academy of Sciences}, vol.~74,
  no.~5, pp.~1800--1801, 1977.

\bibitem{zucker201170+}
I.~Zucker, ``70+ years of the {W}atson integrals,'' {\em Journal of Statistical
  Physics}, vol.~145, pp.~591--612, 2011.

\end{thebibliography}
\bibliographystyle{ieeetr}

\end{document}